\documentstyle[graphics]{cupconf}

\ifoldfss
\else
  \ifnfssone
    \newmathalphabet{\mathit}
      \addtoversion{normal}{\mathit}{cmr}{m}{it}
      \addtoversion{bold}{\mathit}{cmr}{bx}{it}
    \newmathalphabet{\mathcal}
      \addtoversion{normal}{\mathcal}{cmsy}{m}{n}
    \else
    \ifnfsstwo
    \fi
  \fi
\fi

\def\ie{\mbox{i.e.\ }}
\def\gtrsim{\mathrel{\hbox{\rlap{\hbox{\lower4pt\hbox{$\sim$}}}\hbox{$>$}}}}
\def\lesssim{\mathrel{\hbox{\rlap{\hbox{\lower4pt\hbox{$\sim$}}}\hbox{$<$}}}}

\def\hexnumber#1{\ifcase#1 0\or1\or2\or3\or4\or5\or6\or7\or8\or9\or
 A\or B\or C\or D\or E\or F\fi }

\makeatletter
\ifx\CUP@mtlplain@loaded\undefined
\else
\fi
\makeatother
 \makeatletter
 \ifx\CUP@mtlplain@loaded\undefined
   \font\tenbmi=cmmib10 at 10pt
   \font\sevenbmi=cmmib10 at 7pt
   \font\fivebmi=cmmib10 at 5pt

   \newfam\bmifam
   \textfont\bmifam=\tenbmi
   \scriptfont\bmifam=\sevenbmi
   \scriptscriptfont\bmifam=\fivebmi
   
 \fi
 \makeatother
\ifnfsstwo

\fi
\ifnfssone

\fi
\ifoldfss

\fi

\mathchardef\varLambda="0103

\makeatletter
\ifx\CUP@mtlplain@loaded\undefined
\else
\fi
\makeatother
\makeatletter
\ifx\CUP@mtlplain@loaded\undefined
  \font\tenbms=cmbsy10
  \font\sevenbms=cmbsy10 at 7pt
  \font\fivebms=cmbsy10 at 5pt
  \newfam\bmsfam
  \textfont\bmsfam=\tenbms
  \scriptfont\bmsfam=\sevenbms
  \scriptscriptfont\bmsfam=\fivebms

  \edef\bsy@{\hexnumber\bmsfam}
  \mathchardef\bnabla="0\bsy@72
\fi
\makeatother
\def\eg{{e.g.\ }}
\def\etc{{etc.\ }}
\def\etal{\mbox{\it et al.}}

\title[Compact Objects and Accretion Disks]{Compact Objects and Accretion
Disks}

\author[Roger Blandford {\it \etal\/}]%
{R\ls O\ls G\ls E\ls R\ns B\ls L\ls A\ls N\ls D\ls F\ls O\ls R\ls D$^1$,\ns
 E\ls R\ls I\ls C\ns A\ls G\ls O\ls L$^1$,\\
 A\ls V\ls E\ls R\ls Y\ns B\ls R\ls O\ls D\ls E\ls R\ls I\ls C\ls K$^1$,\ns
 J\ls E\ls R\ls E\ls M\ls Y\ns H\ls E\ls Y\ls L$^2$,\\
 L\ls E\ls O\ls N\ns K\ls O\ls O\ls P\ls M\ls A\ls N\ls S$^1$,\ns
 H\ls E\ls E\ls-W\ls O\ls N\ns L\ls E\ls E$^3$
}

\affiliation{$^1$Theoretical Astrophysics, Caltech, Pasadena, CA 91125, USA\\
$^2$Center for Astrophysics, 60 Garden St., Cambridge, MA 02173, USA\\
$^3$Yonsei University, Seoul, Korea}

\begin{document}
\ifnfssone
\else
  \ifnfsstwo
  \else
    \ifoldfss
      \let\mathcal\cal
      \let\mathrm\rm
      \let\mathsf\sf
    \fi
  \fi
\fi

\maketitle

\begin{abstract}
Recent developments in the spectropolarimetric study of compact objects,
specifically black holes (stellar and massive) and neutron stars are
reviewed. The lectures are organized around five topics: disks, jets, 
outflows, neutron stars and black holes. They emphasize physical mechanisms
and are intended to bridge the gap between the fundamentals of polarimetry 
and the phenomenology of observed cosmic sources of polarized radiation, 
as covered by the other lecturers. There has been considerable recent 
progress in spectropolarimetry from radio through optical frequencies and 
this is producing some unique diagnostics of the physical conditions around 
compact objects. It is argued that there is a great need to develop a 
correspondingly sensitive polarimetric capability at ultraviolet through 
$\gamma$-ray energies.   
\end{abstract}
Spectropolarimetric observations, particularly those at radio and
optical wavelengths, have played an important role in high energy
astrophysics. From the discovery of synchrotron radiation  to the
first good evidence for AGN unification, from the  polarization
patterns in the coherent emission of radio pulsars to the discovery of
variable, linear polarization in the absorption troughs of broad
absorption line quasars, polarization studies often provide the best
and sometimes the only clue we have as to the  geometric disposition
of the emitting elements in these  diverse sources when we cannot
resolve them directly.

These notes summarize lectures delivered by Roger Blandford at the XII
Canary Islands Winter School on Astrophysical Spectropolarimetry.
They are written up with the assistance of Eric Agol (Disks), Leon
Koopmans (Jets), Hee-Won Lee (Outflows), Jeremy Heyl (Neutron Stars)
and Avery Broderick (Black Holes) The lectures were intended to
provide a bridge between the general  physical foundations of
polarimetry and its practical description presented at the school by
Drs.~Landi Degl'Innocenti and Keller and the observationally oriented
lectures of Drs.~Antonucci and Hildebrand. They also make some
important connections  to solar, stellar (especially white dwarf)  and
maser polarimetry as described by Drs.~Stenflo, Mathys and Elitzur,
respectively. They are organised  around five generic astrophysical
sources: disks, jets, winds, neutron stars, and black holes. In each
case  a cursory motivation is provided by summarizing  some relevant
observations and presenting some  of the key issues that polarimetry
can help to resolve.  This is followed by an heuristic discussion of
some relevant physical mechanisms in a manner which, it is hoped, will
allow them to be  applied elsewhere followed by a brief account of how
they have been used so far and some suggestions of possible future
investigations.
\section{Disks}
\subsection {Motivation}
Accretion disks are commonly found when gas, with angular momentum, is
gravitationally attracted towards a central massive body (Frank, King,
\& Raine 1992). First described in the context of Laplace's nebular
hypothesis and first seriously analyzed by L\"ust (1952) they have
been observed around black holes, neutron star and white dwarf
binaries (Shapiro \& Teukolsky 1983), around protostars, and
especially within active galactic nuclei, including quasars
(Blandford, Netzer, \& Woltjer 1990, Krolik 1999). It is this last
type of disk that provides us with much of our most detailed
observations. Although it is not part of my task  to discuss them,
observations of young stellar objects and  cataclysmic variables  are
turning out to be particularly instructive and much of what follows is
informed by the results of these studies.

Disks are planar structures and if their opacity is predominantly
scattering, by either free electrons or dust grains, then the
direction and strength of the polarization tells us about the
orientation and  inclination of the disk as well as the location of
the continuum source.  As we shall discuss, (cf also Antonucci,
Hildebrand, these proceedings),  most astrophysical disks are
associated with jets or bipolar outflows  and when these can be
resolved, they may represent the projected rotation  axis of the inner
disk. (As we shall also see, disks can be warped and  this axis can
change with radius and, in the case of AGN, it may be quite  different
from the axis of the host galaxy.)

We wish to use polarization observations to determine what disks are
really like. Unfortunately, the current observational capability  is
limited. Polarimetry in the radio, the near infrared and the optical
regions of the spectrum is really quite good  by astronomical,
(although not solar),  standards.  Optically, spectropolarimetry has
been performed at the 0.001 level down to $R\sim18$. Measurement in
the mid and far infrared is more of a challenge, though one that has
been met in the  far infrared, (Hildebrand, these proceedings).

However, to understand the inner disk we need ultraviolet and  X-ray
polarimetry. The former was carried out for a while on bright quasars
using  the HST Faint Object Camera, as we shall describe in section 3
below, although this has proved to be a little controversial.  X-ray
polarimetry has really only been accomplished successfully  on a few
bright sources (M\'eszar\'os \etal 1988). There are plans to fly a
more sensitive polarimeter on Spectrum-X. It will become clear that
there is a very strong scientific case to be made to develop  X-ray
polarimetry. There is also a strong incentive to  develop a
$\gamma$-ray polarimetric capability though, here, the technical
challenges are even greater. In principle, Compton telescopes
operating at $\sim$~MeV energies, record polarimetric information
though, in practice, it has proven to be almost  impossible to extract
this signal from observations taken to date.
\subsection{Observation}
A particularly good example of an AGN accretion disk can be found in
NGC~4258 (cf Elitzur, these  proceedings). Here water maser
observations reveal a resolved, disk orbiting a 43 million solar mass
black hole (beyond all reasonable doubt). We have believed for a long
while that gas moves radially inwards through this disk as a result of
a hydromagnetic torque that transports angular momentum  outward. The
binding energy that is released by the infalling gas can be radiated
away and this process accounts for the most luminous of quasars and
binary X-ray sources.  It can also be responsible for driving powerful
outflows, as we shall  see. Evidence that disks can extend all the way
down to the  central compact object has been provided by ASCA X-ray
observations of  relativistically-broadened fluorescent iron emission
lines from low luminosity Seyfert and LINER galaxies.  (This
interpretation has been somewhat controversial; observations, with
superior sensitivity and spectral resolution, from XMM-Newton are
therefore eagerly awaited.)

Not all disks are thin. There are good phenomenological reasons  to
suspect that the disks contained in  many Seyfert and LINER galaxies
thicken over some radii to form dense, obscuring torii. Theoretically,
it has been proposed that the inner regions of disks that are supplied
with gas at a rate that  is either much higher or much lower than the
Eddington rate (given by $\dot M_{{\rm Edd}}= L_{{\rm Edd}}/c^2=4\pi
GM/c\kappa$) will thicken because the gas will be unable to cool and
the inflow may even become quasi-spherical.  Similarly, the disk in
NGC~4258 is clearly warped and this is thought to be a common
occurrence. It has even been proposed that radiative torques acting on
disks can turn them over (Pringle 1997).  Polarization observations
offer the opportunity to probe the complex  geometry of these flows.

\begin{figure}[t!]
\resizebox{\hsize}{!}{\includegraphics{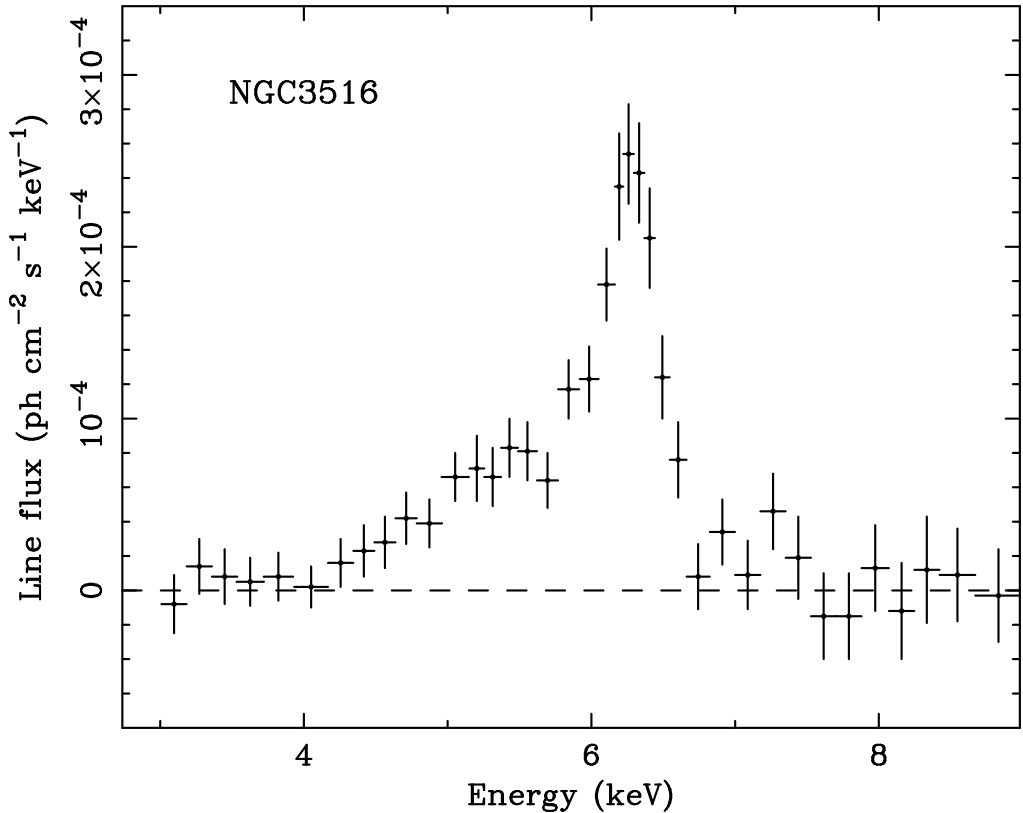}}
\hfill
\parbox[b]{\hsize}{
\label{fig_ngc3516} 
\caption{Broad Fe K$\alpha$ line from NGC 3516 observed
with ASCA by Nandra \etal (1999).}}
\end{figure}

Another example of a subtle arrangement of the emitting elements is
provided  by X-ray observations of the thin disks in Seyfert
galaxies. The current  working model is that only a fraction of the
binding energy released by the accreting gas is radiated from the disk
photosphere (with an effective temperature in the keV range). The
remainder is dissipated in  a hot corona, presumably as a consequence
of magnetic flaring followed by  reconnection and hydromagnetic wave
damping. The heated electrons can then scatter the escaping soft
photons and, as the corona is probably Thomson thick, this will lead
to a power-law tail in the hard X-ray spectrum.  However, roughly half
of these photons will strike the disk where they can be absorbed if
they have low  energy and suffer inelastic Compton recoil loss at high
energy. The reflected  spectrum will therefore be convex and be
imprinted with line features, most famously, the 6.4~keV $K\alpha$
line of Fe (Fabian \etal 2000 and  references therein,
Fig.~(\thefigure)). On this basis, it has been argued that the widths
of lines in  Seyfert and LINER galaxies like MCG 6-30-15 imply that
the central black hole  is rapidly spinning.  However, the details
depend upon the relative location of the region where the
Comptonisation is taking place and the region of the disk responsible
for most of the reflection. There are two ways  to test these models
using more detailed observations. The first is to use a technique
called reverberation mapping (\eg Young \& Reynolds 2000). This
requires monitoring the  variation of the line and the continuum
simultaneously and deriving the lag in the variation of the former in
response to the latter. This  tells us about the geometry and size of
the line-emitting  region. The second technique is to use polarization
observations which  will test the geometry.
\subsection{Physical Processes}
We now summarize some relevant physical processes. A good general
reference is Rybicki \& Lightman (1979).
\subsubsection{Thomson Scattering}
Classical Thomson scattering is strongly polarizing. A free electron
can be considered as a Larmor dipole driven by the electric field  of
the incident wave (with polarization vector $\vec e$).  The scattered
power into polarization is $\vec e'$ is given by
\begin{equation}
{d\sigma\over d\Omega'}=r_e^2(\vec e\cdot\vec e')^2\,,
\end{equation}
where $r_e=e^2/m_ec^2=2.82\times10^{-13}$~cm is the classical electron
radius.  Note that when the scattering angle is $\phi=90^\circ$, the
radiation is  100 percent polarized.

Averaging over incident polarization and summing over final
polarization states gives the familiar differential cross section
\begin{equation}
{d\sigma\over d\Omega'}={1\over2}r_e^2(1+\cos^2\phi)\,.
\end{equation}
Integrating over  solid angle gives the total Thomson cross section
\begin{equation}
\sigma_T={8\pi\over3}r_e^2=6.65\times10^{-25}{\rm cm}^2\,.
\end{equation}
\subsubsection{Compton scattering}
At X-ray and $\gamma$-ray energy, we must take account of the electron
recoil. Conserving energy and linear momentum, we obtain an expression
for the scattered energy $\epsilon'$ in terms of the incident energy
$\epsilon$
\begin{equation}
\epsilon'={\epsilon\over1+{\epsilon\over m_ec^2}(1-\cos\phi)}\,.
\end{equation}
Averaging over $\phi$ for small values of $\epsilon$,  we obtain the
mean energy shift
\begin{equation}
<\Delta\epsilon>=-{\epsilon^2\over m_ec^2}\,.
\end{equation}
The Thomson cross section must be replaced by the Klein-Nishina cross
section
\begin{equation}
{d\sigma\over
d\Omega'}={1\over2}r_e^2\left({\epsilon'\over\epsilon}\right)^2
\left[{\epsilon'\over\epsilon}+{\epsilon\over\epsilon'}-\sin^2\phi\right]\,,
\end{equation}
which emphasizes forward over backward scattering. High energy
scattering is nearly as strongly polarizing as Thomson scattering.
Averaging over $\phi$ again for small angle, we obtain
\begin{equation}
\sigma=\sigma_T\left(1-{2\epsilon\over m_ec^2}+\dots\right)\,.
\end{equation}
For a large photon energy, $\epsilon>>m_ec^2$, we have
\begin{equation}
\sigma\sim{3\sigma_T\over8\epsilon} \left[\log\left({2\epsilon\over
m_ec^2}\right)+{1\over2}\right]\,.
\end{equation}
The rate at which the photons heat the electrons through Compton
recoil is therefore given for $\epsilon<<m_ec^2$ by
\begin{equation}
W_+=n_e\sigma_Tc\int d\epsilon N(\epsilon) {\epsilon^2\over
m_ec^2}=n_e\sigma_TcU{<\epsilon>\over m_ec^2}\,,
\label{wplus}
\end{equation}
where $N(\epsilon)$ is the photon number density per unit energy,
$U=\int d\epsilon\epsilon N$ is the photon energy  density and $<>$
should be interpreted as  an energy density-weighted photon energy.

These expressions describe the energy loss in the initial rest frame
of the scattering electron. However when the plasma is hot  the
electron will be moving and this will cause the photon to experience a
Doppler shift. To $O(v/c)$, blue shifts balance redshifts and there is
no  net energy change when the electron distribution is
isotropic. However, it is apparent what there will be a net energy
gain to $O(v/c)^2=O(kT/m_ec^2)$,  because the rate of approaching
collisions will exceed the rate of receding collisions. We can
therefore express the net rate of energy loss by the electrons in
terms due to the Doppler shift as
\begin{equation}
W_-=n_e\sigma_T c U {xkT\over m_ec^2}\,,
\label{wminus}
\end{equation}
where $x$ is a number that we can fix by observing that there should
be energy balance, $W_+=W_-$, when we use a  dilute black body of
temperature $T$
\begin{equation}
N(\epsilon)\propto\epsilon^2\exp[-\epsilon/kT]\,.
\label{dilute}
\end{equation}
(It is necessary to use a dilute black body to avoid having to
consider nonlinear, induced Compton scattering.) As $<\epsilon>=4kT$,
we deduce that $x=4$.

If this were the dominant physical process,  then the equilibrium
electron temperature in a given radiation field would be
\begin{equation}
T_c={<\epsilon>\over4k}\,.
\end{equation}
However, accretion disk corona are probably heated through
reconnection and hydromagnetic turbulence, and the temperature is
probably  quite non-uniform and hard to predict in detail.  An
additional complication is that  the 
thermalization timescales are
actually quite long  compared with the disk dynamical timescale and so
the plasma is likely  to have a significant suprathermal component
(Gierlinski  \etal 1999).

More generally, we can deduce the form of the  kinetic equation for
the photon distribution. As the individual  photon energy shifts are
small, this will have the form of a modified  diffusion equation in
energy space. However, as the scattering angles are typically large,
we cannot regard this as a diffusion in momentum space.  We therefore
just consider  an isotropic radiation field to bring out some
principles (although this approximation is inappropriate for computing
polarization,  since an isotropic radiation field creates zero net
polarization upon scattering).  As Compton scattering conserves the
number density of photons the  equation must have the form
\begin{equation}
{\partial N\over\partial t}=-{\partial F\over\partial\epsilon}\,,
\end{equation}
where $F$ is the flux of photons in energy space. Now for a dilute
black body, Eq.~(\ref{dilute}), $F$ will be linear in $N$, as long as
we can ignore induced scattering, and, as it represents a diffusion,
$F$ will contain the first derivative of $N$ with respect to
energy. As $F$ must vanish for a dilute black body it must have the
form
\begin{equation}
F(\epsilon)=-g(\epsilon)\left({\partial(N/\epsilon^2)\over\partial\epsilon}+
{N\over\epsilon^2kT}\right)\,,
\label{phflux}
\end{equation}
for some function $g(\epsilon)$.  We next multiply this equation by
$\epsilon$ and integrate over energy and use either Eq.~(\ref{wplus})
or Eq.~(\ref{wminus}) to identify the function
$g(\epsilon)=n_e\sigma_Tc\epsilon^4kT/m_ec^2$.  The resulting
(Fokker-Planck) equation is,
\begin{equation}
{\partial N\over\partial t}={n_e\sigma_T\over m_ec}
{\partial\over\partial\epsilon}\epsilon^2\left[\epsilon^2 kT
\left({\partial(N/\epsilon^2)\over\partial\epsilon}\right)+N\right]\,.
\label{kompaneets}
\end{equation}
This formalism describes the behavior of electrons interacting  with
a dilute gas of photons inside  a box with reflecting walls. If we
need to take account of induced scattering processes, then
Eq.~(\ref{phflux}) must be modified so  $F=0$ when $N(\epsilon)$ has
the Planck form,
\begin{equation}
N(\epsilon)={8\pi\epsilon^2\over
h^3c^3}\left[\exp(\epsilon/kT)-1\right]^{-1}\,.
\end{equation}
The result is that $N\rightarrow N(1+N)$, in the second term in
brackets in Eq.~(\ref{kompaneets}) which is then known as the
Kompaneets equation and is central to discussions of the transfer of
radiation through hot plasmas.

Clearly the radiation as described by this isotropic formalism will be
unpolarized.  In order to describe the polarization of a cosmic
source, we must tackle the radiative transfer. There are three
approaches that have commonly been followed.
\begin{enumerate}
\item{\it Escape Probability Formalism}  This is the simplest
approach. We add a term to the right hand side of
Eq.~(\ref{kompaneets})
\begin{equation}
-{Nc\over R(1+\tau)}\,.
\end{equation}
The extra factor $1+\tau$ takes into account the impeded  photon
escape when the Thomson depth $\tau$ exceeds unity.  This approach,
which is most commonly used, although instructive as far as the
spectrum goes, is not much help when it comes to polarization.
\item{\it Intensity Formalism} Provided that we restrict our attention
to simple shapes -- slabs, spheres \etc -- we can incorporate the
energy space  transport within the equation of radiative transfer
using a scattering kernel. This can then be solved by taking moments
and imposing a closure relation in the standard manner. It is possible
to  include polarization though this leads to quite involved equations.
\item{\it Monte Carlo Formalism}  In many ways the most versatile
method is the same one used  in nuclear reactors for the transport of
neutrons. This is to follow  individual photons, within the scattering
region starting with energies, locations and directions that are
selected according to a prescribed  distribution using random
numbers. Polarization is relatively easy to  handle, and most
polarization is created in the last few scatterings, reducing the
computational burden which occurs at large optical depths.  All of
this is quite straightforward in principle, though, in practice,
Monte Carlo simulations are  quite an art as a variety of ingenious
tricks have to be used to reduce the variance with a finite amount of
computer time.
\end{enumerate}
\subsubsection{Dust Scattering}
Dust scattering is more complex  than Thomson scattering as the cross
section depends on wavelength, grain  size, and grain composition (cf
Hildebrand, these proceedings).   In the limit when the wavelength is
much greater than the grain size,  Rayleigh scattering applies, which
has the same angular cross section as Thomson scattering, but scales
as $\lambda^{-4}$, a fact which has been used to distinguish  electron
scattering from dust scattering in some Seyfert 2 galaxies.   For a
range of different sizes and compositions, it is impossible to
express the dust scattering cross section in a simple formula, but
extensive numerical calculations have been carried out by, e.g.,
Draine \& Lee (1984), Zubko \& Laor (2000).
\subsubsection{Faraday Rotation}
The next relevant physical process is Faraday rotation. When
electromagnetic wave modes propagate through a plasma, their phase
velocities will be changed from $c$. To lowest order, the eigenmodes
are circularly polarized with phase velocity difference
\begin{equation}
\Delta V_\phi=2c\frac{\omega_p^2\omega_G}{\omega^3}\cos\alpha\,,
\end{equation}
where $\omega_p=(4\pi n_e e^2/m_e)^{1/2}$ is the electron plasma
frequency, $\omega_G$ is the electron gyro frequency and $\alpha$ is
the angle between the ray and the magnetostatic field. If we decompose
a linear polarized wave into two circularly polarized modes that
propagate through the medium and then recombine the modes after they
leave the medium, then there will be a net rotation of the plane of
polarization through an angle
\begin{equation}
{d\Phi\over d\tau_T}={\omega\over2}\int ds{\Delta V_\phi\over c^2}=0.1
\left({\lambda\over500\,{\rm nm}}\right)^2\left({B_\parallel\over1\,{\rm
G}}\right)\,.
\end{equation}
Polarization observations can tell us as much about the intervening
medium  as about the source.
\subsubsection{Relativistic Radiative Transfer}
Another interesting complication is that special and general
relativistic effects will affect  the transfer of radiation from the
disk to us. This is particularly interesting for line radiation. The
non-relativistic  Doppler shift will broaden the profile of a line
formed at the photosphere of a rotating disk, with the blue wing
coming from the approaching limb and the red  wing from the receding
limb. The gravitational shift (which is not  separated from the
Doppler shift in a general relativistic calculation) will accentuate
the red wing whose extent depends upon how close the  inner edge of
the disk gets to the hole.  A further effect is that rays  will be
deflected by the gravitational pull of the central black hole  so that
an image of an accretion disk observed from near the equatorial plane
would exhibit the back side of the disk. (There are ambitious
proposals to deploy an X-ray interferometer in space which could
exhibit these and other effects.)

\begin{figure}[t!]\label{fig_polimg}
\resizebox{\hsize}{!}{\includegraphics{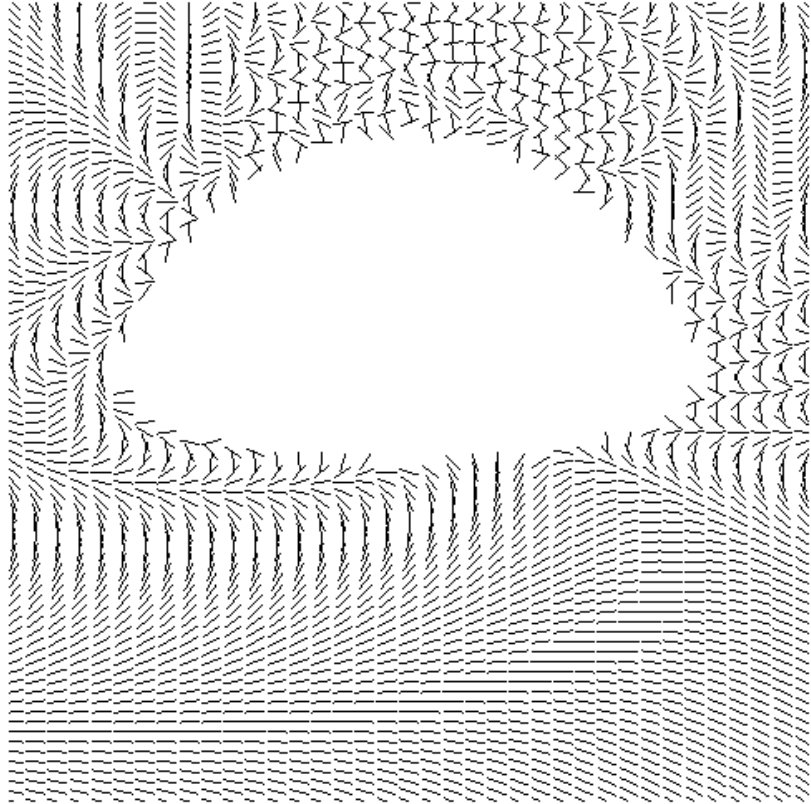}} \hfill
\parbox[b]{\hsize}{
\caption{Polarization angle as a function of position for an
electron-scattering dominated thin accretion disk around a Kerr black
hole (a=0.998) viewed at infinity from an angle of  75$^\circ$.  The
figure is 20$GM/c^2$ in size.}}
\end{figure}

Of more direct relevance to this school,  is the behavior of the
polarization.  There are two main effects. Firstly, aberration changes
the emission angle, and consequently the emitted polarization, in the
rest frame of the orbiting gas.  Secondly, the plane of polarization
is rotated as the ray propagates near the black hole
(Fig.~(\thefigure)).  As we discuss in more detail in \S 5, the
polarization direction is parallel-propagated along  null
geodesics. All of this is straightforward, if somewhat  tedious, to
compute. Specific models for continuum emission have  been computed by
Laor \etal  (1990) and Connors \etal (1980), while relativistic
effects on line polarization were computed by Chen \& Eardley (1991).
It may be possible to use observations  of the rotation of the
polarization direction with wavelength, in a spectral line, or in the
continuum if the wavelength is a measure of the effective radius of
the disk, to  measure the spin of the hole.

\subsection{Interpretation}
Having outlined some of the relevant physical mechanisms,  let us
return to the problem with which we began this section.  ``How much
polarization do we expect from an accretion disk and what do we
observe?'' We can split the problem into two parts -- the escape of
photons emitted in the disk and the behavior of the scattered
photons.  For a given atmosphere, the problem is linear and we can
superpose the  two components. However, if we try to solve
self-consistently for the  ionization and thermal state of the
atmosphere,  the problem becomes nonlinear.
\subsubsection{Electron Scattering} 
Radiative transfer in a pure scattering, plane parallel atmosphere is
a classical problem that was solved for a Thomson scattering kernel,
initially analytically,  by Chandrasekhar (1960) and then in greater
generality by Angel (1969) using a  Monte Carlo approach. The answer
is that the emergent polarization varies with inclination, having a
value $p=0.12$ when the atmosphere is viewed horizontally and $p=0.02$
when viewed at the most probable inclination of $60^\circ$ and, of
course, $p=0$ when viewed normally.  Real disk atmospheres also have
an absorptive opacity and this will  reduce (or increase) the emergent
polarization significantly (Hubeny \etal 2000).  In addition, we now
believe that accretion disks are strongly magnetized. The rationale
for this is that ionized accretion disks are known to be unstable to
developing strong internal magnetic fields with interior magnetic
pressures estimated to be $\sim1-10$~percent of the gas pressure. This
magnetic  field will surely be carried out beyond the photosphere and
into the coronae (discussed above) by buoyancy forces.  Furthermore,
magnetic pressure is likely to dominate gas pressure in an
accretion disk corona, just like in the solar corona.

When we consider the specific parameters appropriate to  observed
disks, we find that thermal emission should be unpolarized based on
the following argument. Consider a given disk annulus,  the radiation
pressure at the photosphere, $aT^4/3$, is smaller than the gas
pressure within the disk.  Because observations at a given thermal
wavelength peak near $\lambda \sim hc/kT$, we can derive a  lower
limit on the  magnetic field strength $B \gtrsim 10^2(\lambda/500\,{\rm
nm})^{-2}\,{\rm G}$.  Since photons traverse differing optical depths after
their last scattering and the magnetic field will likely have
significant inhomogeneities,  any polarization will be erased by the
Faraday rotation  $\langle d\Phi/d\tau_T\rangle \gtrsim 10$, {\it
independent} of the  wavelength observed or the physical size of the
disk (Agol \& Blaes 1996).

Thus, it should come as no surprise that AGN disks are generally only
polarized by a small amount (Antonucci, this volume). Furthermore, the
polarization that is observed  may be imprinted extrinsically.
Purported rises in  polarization below $912\,\AA$ in a few quasars
observed with HST  contradict old theoretical predictions and stand as
a challenge to disk theory (Koratkar \& Blaes 1999).

\begin{figure}[t!]\label{fig_ismfig6}
\resizebox{\hsize}{!}{\includegraphics{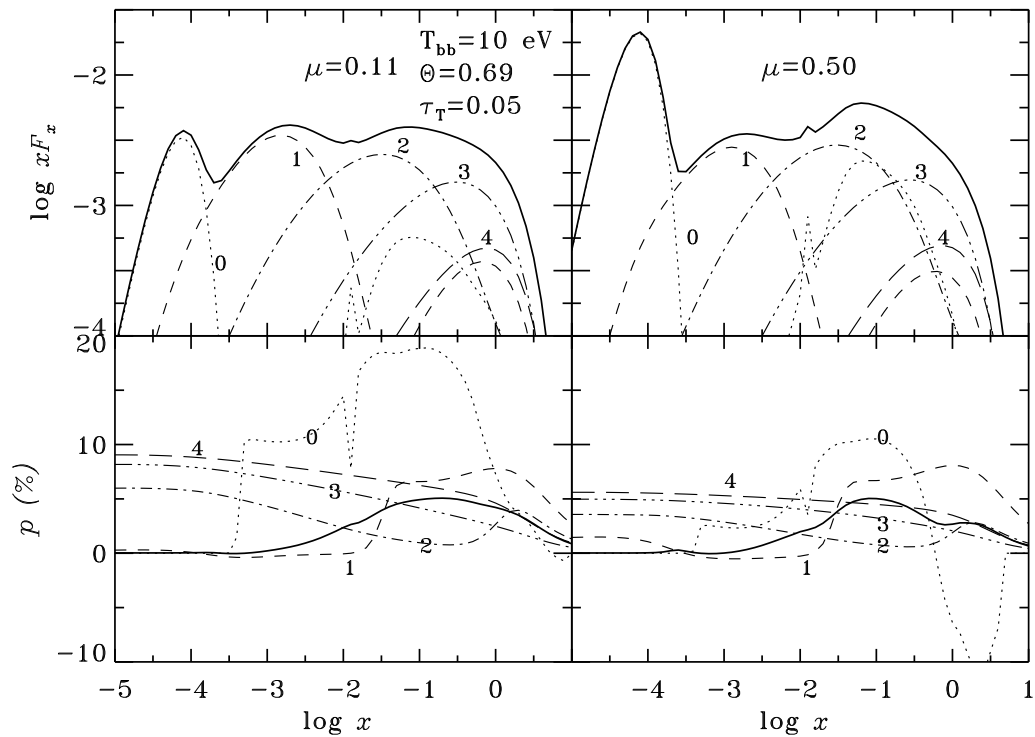}} \hfill
\parbox[b]{\hsize}{
\caption{Spectrum and polarization versus frequency ($x=h\nu/m_ec^2$)
of hot, plane-parallel corona ($T_e =0.11\,m_ec^2$, $\tau_T=0.05$) above
a cold disk viewed at two inclination angles, $\mu = \cos i = 0.11$
and $0.5$.  The numbers label various scattering orders for
reflection, while the solid lines show the total spectrum and
polarization (from Poutanen \& Svensson 1996). No relativistic
propagation effects have been included.}}
\end{figure}

Turning to X-ray wavelengths, where non-thermal emission means that we
might ignore Faraday rotation, but cannot ignore  reflection,
predictions of the emergent polarization under a variety of models
have been presented by  Matt, Fabian, \& Ross (1993) and Poutanen \&
Svensson (1996) and references therein,  as shown in
Fig.~(\thefigure).  The rather flat X-ray spectrum is created
by Compton scattering of thermal emission from the accretion disk,
which is partly due to absorbed X-rays.  The electron-scattering
reflection  feature is suppressed at low energies, $\lesssim 8$ keV, by
X-ray  bound-free absorption opacity, and at high energies, $\gtrsim
100$ keV,  reduced by electron recoil; consequently, the largest X-ray
polarizations should lie between these energies.  The magnitude of the
polarization may depend upon the exact geometry and placement of the
coronal emission regions, an unexplored problem.

\subsubsection{Dusty Disks}

The outer parts of accretion disks may be cool enough ($T<1800$ K) to
be inhabited by dust grains.   If the dusty disk drives a wind, or is
inflated or warped, then dust will scatter the light from the  inner
disk, imprinting a polarization signature from the infrared to
ultraviolet.  It may in practice be  quite difficult to distinguish
between a dusty disk, torus, or outflow using polarization, as the
level of polarization expected is quite small, $\sim 1$\%, and depends
on the details of the dust model,  e.g. K\"onigl \& Kartje (1994).  In
addition, dust extinction can create polarization if the grains,
charged by collisions with ions, are aligned by magnetic fields,
inducing polarization at the  percent level as well.

\subsection{Summary}
\begin{itemize}
\item Disks are commonly found in accreting systems.
\item Model accretion disks can create strong  polarization both in
transmission  and in reflection, throughout the electromagnetic
spectrum.
\item Measurement of the variation of linear polarization with
wavelength can, in principle, reveal a lot about the disk structure
and the  location of coronal emission sites.
\item However, the situation is, in practice, more complex,
particularly at optical wavelengths, where external illumination,
warping, and  especially Faraday rotation are likely to be very
important.
\item Even when polarization cannot be measured, its effects are so
large under conditions of strong electron scattering that radiative
transfer calculations should include polarization.
\item Monte Carlo techniques are well-suited for  computing
polarization in a given model.
\item There is a very strong case for developing X-ray polarimetry.
\end{itemize}
\section{Jets}
\subsection{Motivation}
Jets, or more generally bipolar outflows, are also surprisingly
common.  They have been studied in association with active galactic
nuclei (AGN), binary X-ray sources, young stellar objects, novae and
so on.  Jet speeds are typically a few times the escape velocity from
the central object; in the case of black holes, bulk Lorentz factors
of $\gamma\sim10$ are inferred.  (Gamma ray bursts may also produce
jets with Lorentz factors $\gamma\sim300$.)  Jets are so common that
it has been speculated that they may be an essential concomitant of
accretion flow -- the channels through which the liberated angular
momentum and perhaps also much of the energy leave the system. The
challenge to the astrophysicist is to explain how jets are powered and
collimated. However, even after decades of work, major theoretical and
observational questions about their origin, collimation and even their
constituents still remain.

There are, generically, two proposed origins for the jet power: the
central object (black hole, neutron star or protostar) and the
accretion disk (e.g. Blandford \etal 1990). In both cases,  the
energy derives from differential rotation. For example in the case of
a Keplerian disk that extends down to the surface of a non-rotating,
unmagnetized star, as much energy is released in the boundary layer as
in the disk. An extreme case is presented by the Crab (Weisskopf \etal
2000) and Vela (Helfand, Gotthelf \& Halpern 2001) pulsars which
exhibit prominent jets without there being any accretion
disk, presumably.

One of the best observational approaches to investigate the mechanisms 
which produce jets is to determine the jet composition at 
radii where they can be observed directly. In the
case of black-hole jets, the plasma is likely to be electron-ion if
the jet originates from a disk, or electron-positron if it derives
from the black hole.  Polarization observations have been prominent in
attempts to distinguish between these two possibilities. Most
contemporary explanations of the collimation invoke strong magnetic
fields, though in most cases, the argument for magnetic collimation is
a theoretical one, based upon eliminating the alternatives. One
exception to this is the bipolar outflow associated with young stellar
objects (YSOs) where polarization observations strongly support the
notion that the magnetic field is dynamically important (Akeson \&
Carlstrom 1999).

\begin{figure}[t!]\label{fig_M87jet}
\resizebox{\hsize}{!}{\includegraphics{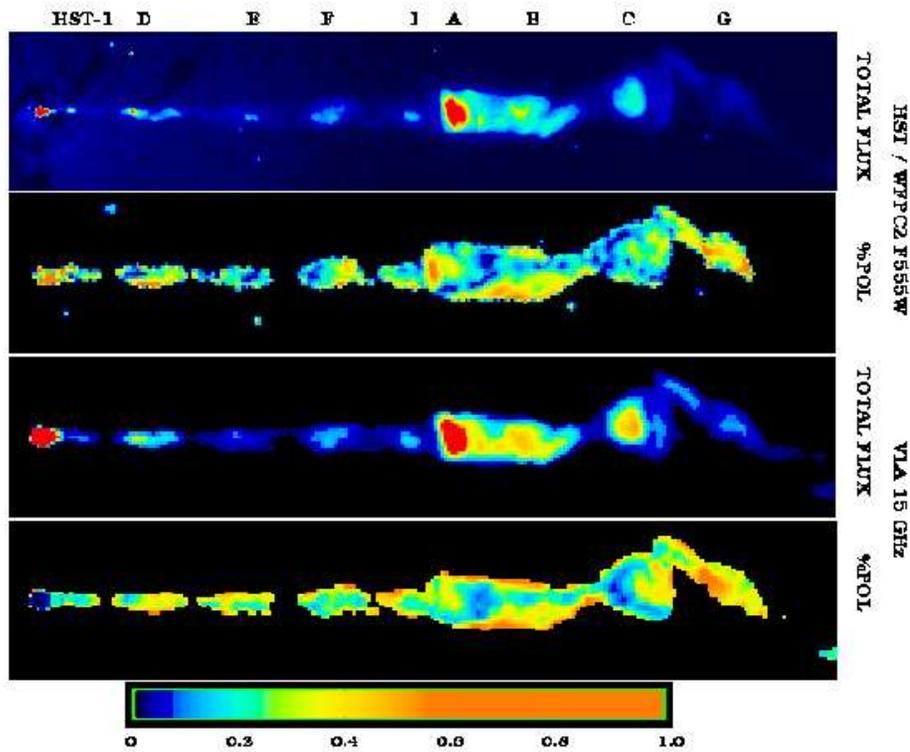}} \hfill
\parbox[b]{\hsize}{
\caption{False-color representations of the total intensity and
polarization of the M87 Jet in the optical (HST F555W, top two panels)
and radio (VLA 14.5 GHz, bottom two panels).  The HST observations
were carried out in May 1995, while the VLA observations were done in
February 1994.  All maps were rotated so that the jet is along the
x-axis, and are convolved to $0.23''$ resolution (from Perlman et
al. 1999).}}
\end{figure}

\subsection{Observations}

The first jet observed was that in the nucleus of the elliptical
galaxy M87 in the Virgo cluster. The modern representation is given in
Fig.~(\thefigure) (Perlman \etal 1999). In this case, the jet
emerges from no more than 100 times the gravitational radius of the
central hole ($M=3\times10^9$~M$_\odot$; $GM/c^2= 4\times10^{14}$~cm)
and propagates radially outward for a distance
$\sim3\times10^{22}$~cm, seven orders of magnitude larger. The jet,
however, is neither homogeneous nor smooth. Superimposed upon an
overall decrease in surface brightness as the jets expand away from the 
central hole are strong side to side variations and bright features. These 
may reflect a time-dependence at
the jet origin or independent, local instabilities. At small radii,
the M87 jet is strongly one-sided, and this is generally attributed to
Doppler beaming and adduced as evidence for relativistic outflow.

\begin{figure}[t!]\label{fig_pictorA}
\resizebox{\hsize}{!}{\includegraphics{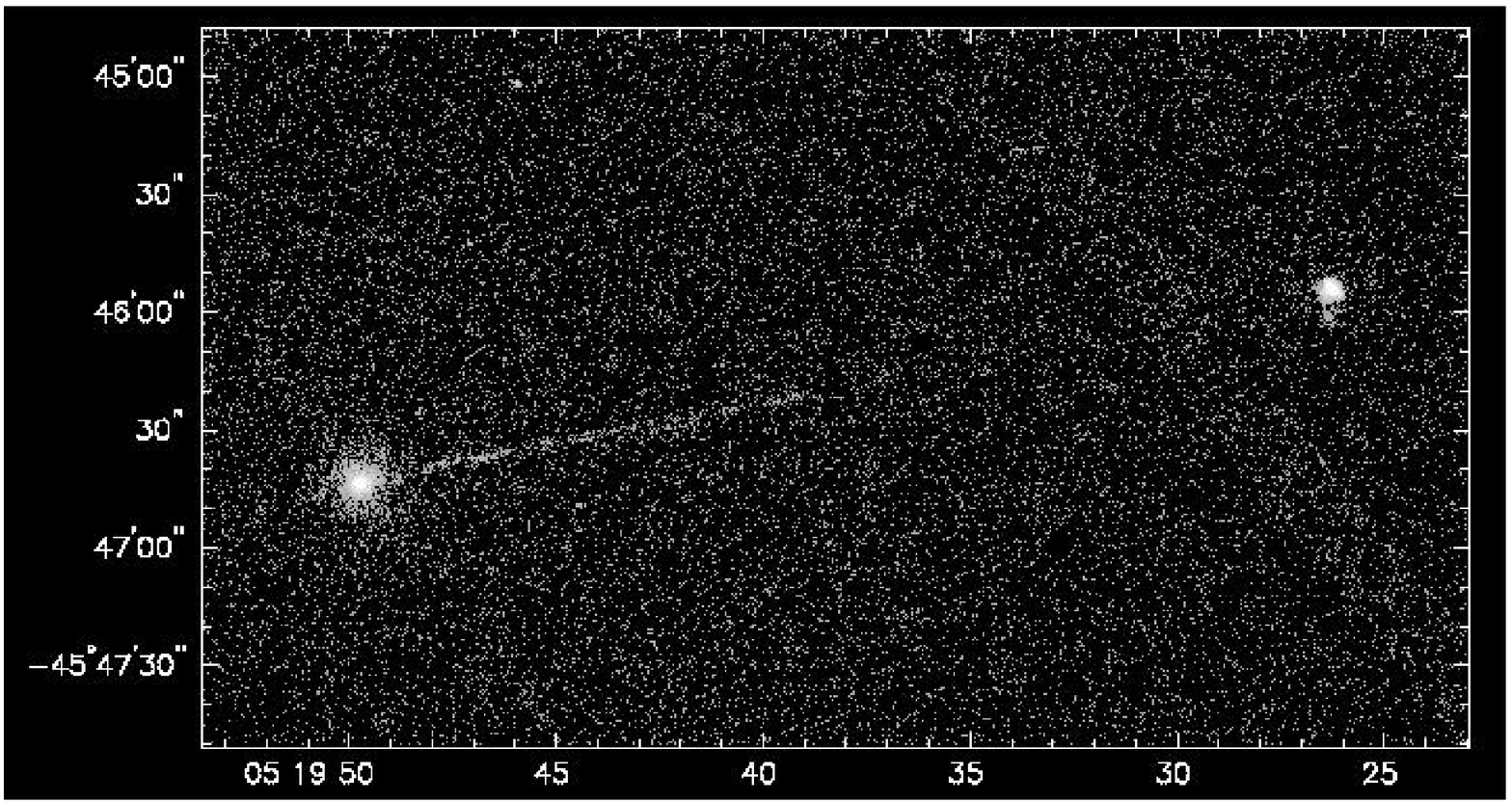}} \hfill
\parbox[b]{\hsize}{
\caption{Gray-scale representation of the full-resolution Chandra
image of the nucleus, jet and western hot-spot of Pictor A (from
Wilson \etal 2001).}}
\end{figure}

Another good example of an extragalactic jet is Pictor A, which has
recently been observed at X-ray energies using Chandra
(Fig.~(\thefigure)); Wilson, Young \& Shopbell, 2001). It shows
strikingly efficient collimation and powerful emission from the
western hot spot, which is also prominent at optical and radio
wavelengths.  The Galactic source GRS 1915+105 shows mild
superluminal motion which is interpreted in terms of moving features,
probably internal shocks, with space velocities $\sim0.9c$
(e.g. Mirabel \& Rodriguez 1994).  This is mild compared with most
extragalactic compact jets where the speeds are more typically
$\sim0.99\,c$, corresponding to bulk Lorentz factors $\gamma\sim7$.
Indeed, there are observational indications that much larger bulk
Lorentz factors are produced in extragalactic jets. Independent
evidence that these jets are relativistic comes from $\gamma$-ray
observations, principally done with the EGRET detector on Compton
Gamma Ray Observatory. These showed that those jets that are beamed
towards us, and which are collectively known as blazars, are often
powerful $\gamma$-ray sources. The total electromagnetic spectrum of
blazars (and similar sources) comprises a broad band synchrotron
radiation spectrum extending from low radio frequencies to an upper
frequency between optical and X-ray wavelengths. The same electrons
are responsible for an inverse Compton component that can extend up to
TeV energies (with variability times as short as $\sim30$~min). (We
can only observe TeV emission from relatively local sources because
TeV photons from cosmologically distant sources will be absorbed
through pair production on the intergalactic infrared background.)
The fastest jets may well be associated with $\gamma$-ray bursts, if
they are indeed beamed, for which speeds of $\sim0.999995c$ have been
inferred.

\subsection{Physical Processes}

\subsubsection{Synchrotron Radiation}

As described in Dr. Landi Degli'Innocenti's contribution (for more
details see also Rybicki \& Lightman 1979), the synchrotron power
radiated by an ultra-relativistic electron with energy $\gamma m_ec^2$
in a field of strength $B$ is given by
\begin{equation}
	P={4\over3}\gamma^2\sigma_TcU_B \,,  \label{synpower}
\end{equation} 
where $U_B=B^2/8\pi$ is the magnetic energy density and we have
averaged over pitch angle.  The corresponding radiative cooling time is
\begin{equation}
	t_S={5\times10^8\over B^2\gamma}\,{\rm s}\,,
\end{equation}
where $B$ is measured in Gauss.  The characteristic frequency radiated
is
\begin{equation}
	\nu_c=\gamma^2 B \,{\rm MHz}\,.
\end{equation}
The polarization of single particle emission varies from $2/3$ for
$\nu\ll\nu_c$ to 1 for $\nu\gg\nu_c$. The electric vector is
perpendicular to the projected magnetic field direction. Averaging
over a power law distribution of relativistic electrons,
$dN/d\gamma=K\,\gamma^{-s}$, it can be easily shown that the observed
intensity is
\begin{equation}
	I_\nu\propto K\,B^{1+\alpha}\nu^{-\alpha}\,,
\label{optthin}
\end{equation}
where the spectral index $\alpha=(s-1)/2$. The net degree of linear
polarization is
\begin{equation}
	{Q\over I}={s+1\over s+7/3}\,.
\end{equation}
Synchrotron radiation is also naturally circular polarized to an extent
\begin{equation}
	{V\over I}\sim 3/\gamma\,,
\end{equation}
dependent upon the detailed angular distribution function and  the
viewing angle.

An important issue for what follows is the viability of a synchrotron
maser. This can be shown to be impossible for ultra-relativistic
emission in vacuo. Essentially, in order for a maser to operate, it is
necessary that there be a population inversion and that the emissivity
at a given frequency decreases sufficiently rapidly with increasing
energy and this does not happen with regular synchrotron emission.
(It can however arise when relativistic electrons emit synchrotron
radiation in a plasma, although in practice the conditions for this to
occur are rather restrictive.  This is known as the Razin effect.)

Of more relevance is what happens when the electrons are no more than
mildly relativistic. The emission is then confined to a series of
harmonics of the fundamental gyro frequency $\omega_G/\gamma$. As the
central frequency of a harmonic decreases with increasing energy,
there are frequencies and directions where maser action is possible,
given a population inversion. Cyclotron masers are likely to be highly
polarized. The polarization from an electron-ion plasma will be
elliptical; that from an equal pair plasma will be purely linear.

When the brightness temperature $T_B=I_\nu c^2/2k\nu^2$ of a
synchrotron source approaches the kinetic temperature of the emitting
electrons, $T_{\rm k}\sim\gamma m_ec^2/3k\propto(\nu/B)^{1/2}$, the
radiation will be absorbed. The optically thick radiation from a
source will have a brightness temperature that is limited to this
value and so the optically thick equivalent of Eq.~(\ref{optthin}) is
\begin{equation}
	I_\nu\propto\nu^{5/2}B^{-1/2}\,.
\end{equation}
When we consider the linear polarization of a self-absorbed source, we
observe that the electrons emitting in the field-perpendicular
polarization at a given frequency will have slightly lower energies
than those emitting in the field-parallel polarization. Therefore the
brightness temperature of the field parallel emission will be slightly
larger than that of the field perpendicular emission. The degree of linear
polarization from a power-law relativistic electron distribution
function can be computed to be
\begin{equation}	
	{Q\over I}={-3 \over 6\,s+13}\,.
\end{equation}
An example is the supermassive black hole candidate in the Galactic
center, Sagittarius~A$^{*}$, which has no linear polarization up to
frequencies of 86 GHz (cf Hildebrand, these proceedings),  but shows
surprisingly strong circular polarization (e.g. Bower 2000).  The
degree of CP increases sharply with frequency. Whereas both
advection-dominated-accretion-flow (ADAF) models and
accretion-disk-powered-jet models can account for the spectrum from
centimeter wavelengths to X-rays of Sagittarius~A$^{*}$, they have
distinct polarization characteristics which might in the near future
be able to distinguish between the two models.

\subsubsection{Inverse Compton Scattering}

Inverse Compton (or more properly Thomson) scattering, in which a
highly energetic particle transfers momentum to a low-energy photon,
is very similar to synchrotron radiation. If the radiation field is
isotropic, then the power is given by Eq.~(\ref{synpower}) with the
magnetic energy density $U_B$ replaced by the radiation energy
density, $U_{{\rm rad}}$. As photons are conserved in Thomson
scattering, the mean photon frequency is boosted by an average factor
\begin{equation}
	{\nu'\over\nu}={4\over3}\gamma^2 \,,
\end{equation}
where $\nu'$ is the scattered frequency and $\nu$ is the incident
frequency.  One power of $\gamma$ arises from the Lorentz
transformation into the electron rest frame; the second comes from the
scattering back into the original frame. The polarization observed
will be generically be $\sim1/\gamma$ unless the incident radiation
field is both highly anisotropic and polarized, in which case a
strongly polarized scattered spectrum can be emitted. This can arise
when, for example, radiation is scattered into a beam by a warped
disk. Similarly, circular polarization in the incident radiation will
be partly retained in the scattered radiation.

\subsubsection{Inverse Compton Limit}

The comparison of synchrotron radiation and inverse Compton scattering
leads to what, for historical reasons, is called the inverse Compton
limit (Kellermann \& Pauliny-Toth 1969). The way the argument is
traditionally expressed is that the ratio of the Compton power
radiated by an electron to the synchrotron power can be written as
\begin{equation}
	{L_{C^{-1}}\over L_S}={U_S\over U_B}\propto{\nu_S^3T_B\over
 	B^2} \propto T^5\nu_S \,,
\end{equation}
in obvious notation, and where we have assumed that the source is
self-absorbed at the observing frequency. If we set $\nu_S\sim
1-10$~GHz, then the brightness temperature is limited to
$T_B\sim2\times10^{12}$~K if this ratio is not to exceed unity.

The original concern was that if the ratio did exceed unity then the
second order Compton scattering would be even greater than the first
order scattering and so on.  Of course this can't go on for too many
orders because the Klein-Nishina limit will limit the scattering.
Nowadays we think we can identify the synchrotron and the inverse
Compton components and so we know their ratio and can deduce the
source brightness temperature which is quite insensitive to its value.

\subsubsection{Kinematics of Bulk Relativistic Motion}

In the case of a relativistic jet, it is often easier to compute the
radiation spectrum in the comoving (primed) frame of the emitting
plasma and then perform a Lorentz boost into the (unprimed) frame of
the observer. The frequency will be boosted by the Doppler factor
(e.g. Blandford \& K\"onigl 1979),
\begin{equation}
	\delta={\nu'\over\nu}={1\over\gamma(1-\beta \cos\theta)} \,,
\end{equation}
where $\theta$ is the scattering angle in the observer frame and
$\beta=v/c$ is the bulk velocity of the plasma in the jet. Note that
for a jet beamed toward us with $\theta<\gamma^{-1}$, the Doppler
factor is $\delta\sim\gamma$. Note also that the rate of change of
observer time $t_{{\rm obs}}$ to proper time $\tau$ satisfies
\begin{equation}
	{d\tau\over dt_{{\rm obs}}}=\delta \,.
\end{equation}
As a consequence, the observed transverse speed of a feature moving
with the jet speed is given by
\begin{equation}
	\beta_{{\rm obs}}= \gamma\beta \sin\theta{d\tau\over dt_{{\rm
	obs}}}= {\beta \sin\theta\over1-\beta \cos\theta} \,.
\end{equation}
This has a maximum value $\gamma\,\beta$ for $\theta=\cos^{-1} \beta$
and consequently the expansion can be ``superluminal'' when
$\beta>0.71$. The kinematics of real jets is undoubtedly more complex
and the space motion of shock features must be distinguished from the
speed of the emitting plasma.  Frequently, observers make the
approximate identification $\beta_{{\rm
obs}}\sim\theta^{-1}\sim\delta\sim(dt/dt_{{\rm obs}}) \sim\gamma$ when
interpreting measurements of compact extragalactic radio sources.

\begin{figure}[t!]\label{fig_boost}
\resizebox{\hsize}{!}{\includegraphics{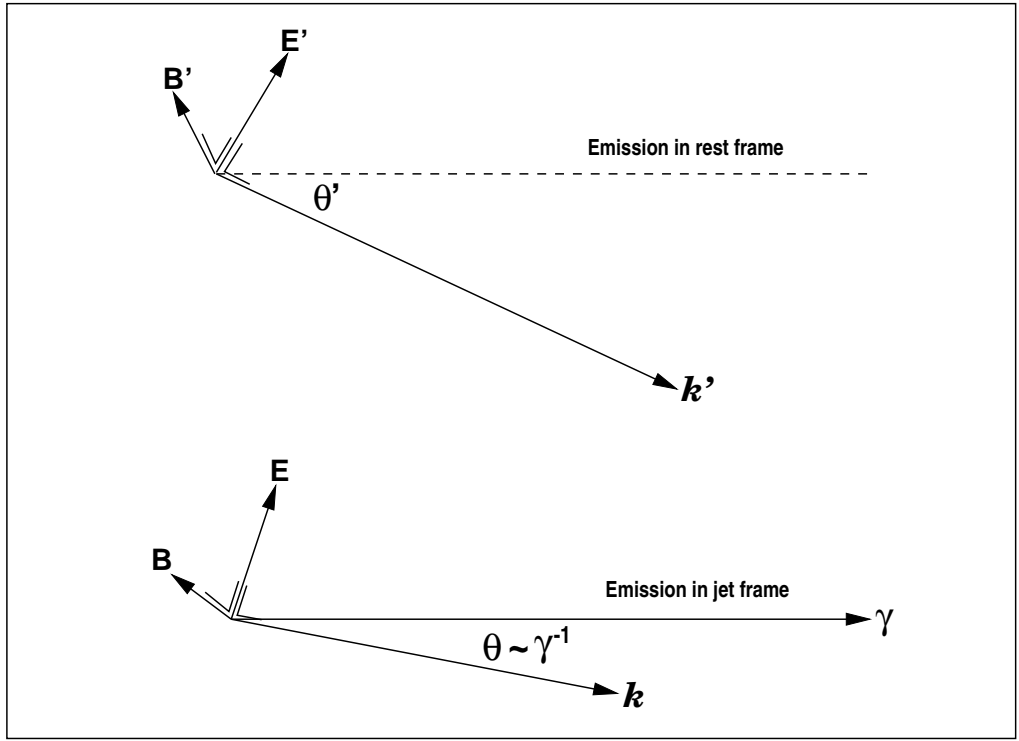}} \hfill
\parbox[b]{\hsize}{
\caption{The Lorentz boost of polarization.}}
\end{figure}

The behavior of polarization under a Lorentz boost is
straightforward.  The k-vector swings forward along the direction of
motion making an angle $\theta'$ with the direction of motion in the
plasma rest frame and an angle $\theta \,$ in the jet frame, where
$\sin\theta'=\delta \sin\theta$.  If we imagine the k-vector as being
rotated in this manner, then the electric and magnetic fields
associated with individual photons will be similarly rotated about a
direction $\vec k\times\vec B$ so that $\vec k$, $\vec E$ and $\vec B$
continue to form an orthogonal triad (Fig.~(\thefigure)).

Because the Planck distribution function is Lorentz scalar, and the 
brightness temperature enters only in the ratio $\nu/T$, it is clear 
that the brightness temperature must transform in the same manner as
frequency.
This implies that if the inverse
Compton limit is applied in the observer frame, the brightness
temperature measured by a radio astronomer can be as high as $\sim
2\times10^{12}\,\delta$~K.  This is particularly germane at this time
because radio astronomers are able to estimate these brightness
temperatures, both directly using ground and orbiting VLBI, and
indirectly by carefully analyzing refractive interstellar
scintillation. Using the limits on the source size derived from these
observations and its relation to the source flux-density (i.e. the
Rayleigh-Jeans equation), one can derive an apparent  surface
brightness temperature and therefore the value of $\delta$, assuming
the intrinsic brightness temperature cannot exceed the inverse Compton
limit significantly. The values that are found require  bulk Lorentz
factors $\gamma\sim\delta\sim 30$ and may be even higher.

\subsubsection{Faraday Conversion}

We discussed Faraday rotation in a cold, non-relativistic plasma in
the last section. We must now consider what happens in an
ultra-relativistic pair plasma. On symmetry grounds, the eigenmodes
must be linearly polarized and are usually labeled ordinary, where the
electric vector along the direction $\vec k\times\vec B$, and
extraordinary, where it is not. If a linearly polarized wave, obliquely 
polarized with
respect to the magnetic field is incident upon the plasma, then it can
be decomposed into ordinary and extraordinary modes that will
propagate with slightly different phase velocities. In this manner,
circular polarization will be created. In other words, there is a
conversion of $U$ to $V$. The sense of circular polarization will be
given by the sign of $(\vec e\cdot\vec B)(\vec e\cdot\vec k\times \vec
B$) and so in order to have a measurable circular polarization from a
cosmic source, it is necessary to have a preferred field
orientation. This phenomenon is known as Faraday conversion (e.g. Jones
\& O'Dell 1977).

Because Faraday conversion is caused by the lowest energy relativistic
electrons, it can serve as a probe of the low energy end of the
electron energy distribution. Faraday conversion is furthermore
proportional to $e^2 B^2/m^2$, i.e. independent from the sign of the
particle's charge. An equal mixture of electrons and positrons  can
therefore produce Faraday conversion, but not rotation. A comparison
of linear and circular polarization might therefore probe the
constituents of the jet (e.g. electron-positron versus electron-ion
pairs).

\subsection{Interpretation}

\subsubsection{Shocks and the Integrated Spectrum}

\begin{figure}[t!]\label{fig_spectra}
\resizebox{\hsize}{!}{\includegraphics{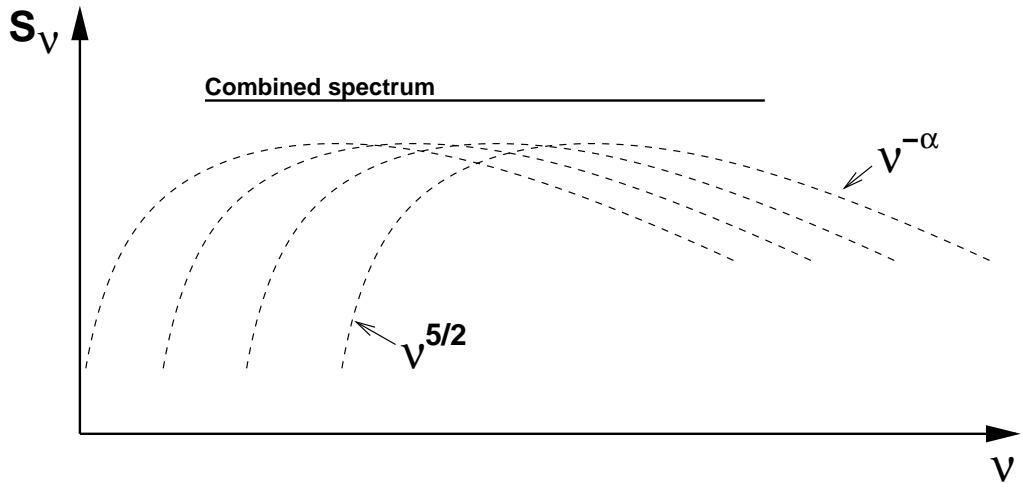}} \hfill
\parbox[b]{\hsize}{
\caption{A superposition of self-absorbed sources peaking at different
frequencies can create a combined `flat' spectrum.}}
\end{figure}

The emitting element in powerful synchrotron jets is thought to be a
relativistic shock wave and a typical source will comprise the
emission from several of them. In the limit, a jet can be thought of
as accelerating relativistic electrons over its length with a field
strength that diminishes with radius. The total flux density observed
at a given frequency is dominated by the emission from the radio
photosphere, where the optical depth is roughly unity. Blazars and
similar sources generally have ``flat'' spectra, \ie $-0.5<\alpha<0.5$
and can be interpreted as the superposition of a series of
self-absorbed sources at successive radii, each peaking at
successively lower frequencies (Fig.~(\thefigure)).

\subsubsection{Jet Composition}

On the basis of the fraction of the jet energy per relativistic
electron responsible for the synchrotron radio emission that we
observe to be radiated on an outflow timescale, it has been
tentatively deduced that jets cannot carry protonic ``baggage'' and so
must comprise electron-positron pairs with a low energy cutoff in the
distribution function (\eg Reynolds \etal 1996).

Wardle \etal (1998) measure a large degree of circular polarization
in 3C279, a source where the linear polarization is also quite high,
and so there cannot be too much normal Faraday rotation. In order to
explain the circular polarization, they have to invoke a large
population of mildly relativistic electrons and positrons. On this
basis they conclude that relativistic jets comprise pair plasmas at
least at the radii where they are directly observed.

\subsubsection{Coherent Emission Mechanisms}

All of this calls into question the fundamental synchrotron
hypothesis.  If the deduced jet powers are unreasonably large then we
should certainly be prepared to consider the possibility that the
radio emission, or at least its compact and variable part, may be due
to a coherent emission mechanism.  This is not unreasonable. After
all, the sun and Jupiter support high brightness coherent emission
under far more docile conditions. Furthermore, a shock front is a very
natural environment in which strong and unstable currents are likely
to be induced and these are commonly observed to radiate coherently in
plasmas. Probably the most likely possibility is coherent cyclotron
emission emerging from very much more compact regions than under the
jet hypothesis. This requires the magnetic field strength to be
hundreds or even thousands of Gauss and so the pressures must be much
larger than in the jets. Coherent cyclotron emission is likely to be
strongly circularly polarized, unless there are equal numbers of
electrons and positrons.

\subsubsection{Microlensing and Refractive Scintillation}

Both microlensing by compact objects in the line-of-sight and
refractive scintillation by density fluctuations in the Galactic
ionized medium can introduce non-intrinsic variability of the compact
structures in jets of extra-galactic radio sources (e.g. knots or
shock fronts). For significant variability to occur, these structures
must have an angular size of the order the Fresnel scale for
scintillation or Einstein radius in the case of microlensing
(e.g. Koopmans \& de Bruyn 2000).  Both scales are typically several
micro-arcseconds. The expected time-scale of variability is determined
by the transverse velocity of the source compared to the scattering
medium (i.e. the compact objects or the Galactic ionized ISM) and is
typically hours to weeks in the case of refractive scintillation or
weeks to months in the case of microlensing.

Whereas polarization is typically little affected by either
scintillation or microlensing (which retains polarization angle), in
both cases only the most compact source structures vary
significantly. If these structures have different polarization degrees
or angles compared with the flux-density weighted average over the
source, the net results will be polarization variability that strongly
correlates with the non-intrinsic source variations.  A correlation
between changes in polarization and non-intrinsic flux variations of
extra-galactic radio sources (for example in intra-day variables
(IDVs), which strongly scintillate) could therefore provide
information on the polarization properties of the most compact
micro-arcsecond scale jet structures, which are impossible to observe
directly in any other way.

There are several other ways that the Galactic ionized ISM can
introduce non-intrinsic variations in the polarization of radio
sources (including radio jets), or even induce circular
polarization. The simplest case is that of extreme scattering events
(ESEs), where large localized overdensities in the Galactic ionized
medium move into the line-of-sight to the radio source.  The enhanced
electron column density increases the Faraday rotation and could
result in an observable change in the polarization angle of the radio
source, as well as in a change of its flux-density due to refractive
lensing. Besides Faraday conversion, which converts linear
polarization to circular polarization, if a strong gradient in
the rotation measure over the source exist, the scintillation patterns of
left and right-hand polarized wavefronts will be slightly
displaced. This will introduce a time-variable circular polarization that is
strongly correlated with the scintillation-induced flux-density
variations (Macquart \& Melrose 2000), but independent of the degree
of linear polarization (which is not the case for Faraday conversion).

\subsubsection{Inverse Compton Scattering}

A major concern in interpreting the inverse Compton X-ray and
$\gamma$-ray observations of blazars is the source of the incident
photons. In the lower power objects that can be observed at TeV
energies, these are thought to be synchrotron photons emitted locally
within the jet. In the higher power quasars where there is a powerful
photoionizing continuum as well as a relativistic jet coming towards
us, most of the incident photons are thought to originate from the
disk and to be scattered into the jet. However, this is not certain
and X-ray (or $\gamma$-ray) polarimetry could validate this, because 
these scattered X-rays of external origin would be highly polarized.

\subsection {Summary}
\begin{itemize}
\item Jets appear to be a common and perhaps even a necessary feature
of accreting or possibly even just simply rotating systems.
\item Despite much observational progress, the fundamental  questions
concerning the origin, composition and collimation  of jets remain
unanswered.
\item Black hole jets are formed ultra-relativistically as required to
account for their superluminal motion, high radio brightness
temperature and prodigious $\gamma$-ray emission.
\item Radio polarimetry of relativistic jets is helping  us to deduce
their composition although coherent emission mechanisms cannot be
ruled out.
\item X-ray and $\gamma$-ray polarimetry could help refine  models of
relativistic jets by probing jets close to their origins.
\end{itemize}
\section{Outflows}
\subsection{Motivation}
Somewhat paradoxically, accreting systems commonly exhibit
outflows. The reason for this behavior is simple. As gas accretes
onto a compact object it must  release its gravitational binding
energy. When this is possible, it will  do so by radiating. However,
this may not be possible when the gas accretes much faster than the
Eddington rate, $\dot M_{{\rm Edd}}=4\pi GM/\kappa_Tc$, the photons
will be trapped by electron scattering and the energy  can only be
carried off by a bulk outflow. Even if the radiation is not trapped,
then it can still drive an outflow if it encounters gas with an
opacity $\kappa>>\kappa_T$, for example with  dust grains or resonance
lines. Either, or more likely both, of these processes are believed to
drive the outflows associated with  Broad Absorption Line Quasars (BALQs)
and this lecture will be primarily about these objects, although
the principles involved are more generally  applicable. The broad
absorption lines by which these quasars are distinguished are
associated with the ultraviolet resonance lines of the common ions.
(Recall that quasars are mostly at high redshift and so these lines
are observed in the visible.) They show absorption troughs, extending
to the blue of the regular, broad emission lines by which quasars are
identified spectroscopically. This is just what is also seen in the
star P Cygni, though the BALQ relative velocities, $\sim0.1c$, which
are what one might escape for gas escaping from  the vicinity of a
black  hole, are much larger than those encountered in  P Cygni
(Weymann \etal 1991).

However, we do not understand the flow of gas around the black hole
and the location of the emission line and the absorption line gas is
still quite controversial.  Models of broad line clouds have been
constructed in which the  gas flows ``in, out, round or about'' and
there may be elements of each of these four kinematic classes in real
sources.  It is probably safe to conclude that the flow is not very
simple; otherwise we would have already understood it using a
technique called ``reverberation mapping''. In this technique it  is
supposed that the lines vary in direct response to the photoionizing
continuum. By monitoring them both, it is possible  to construct a
Greens function response of the emission line gas, This can then be
compared with  the predictions of simple kinematic models. It has been
possible  to use this technique to locate the gas in several
instances.  However, the details of the velocity field remain
controversial.

\begin{figure}[t!]\label{fig_seyfert}
\resizebox{\hsize}{!}{\includegraphics{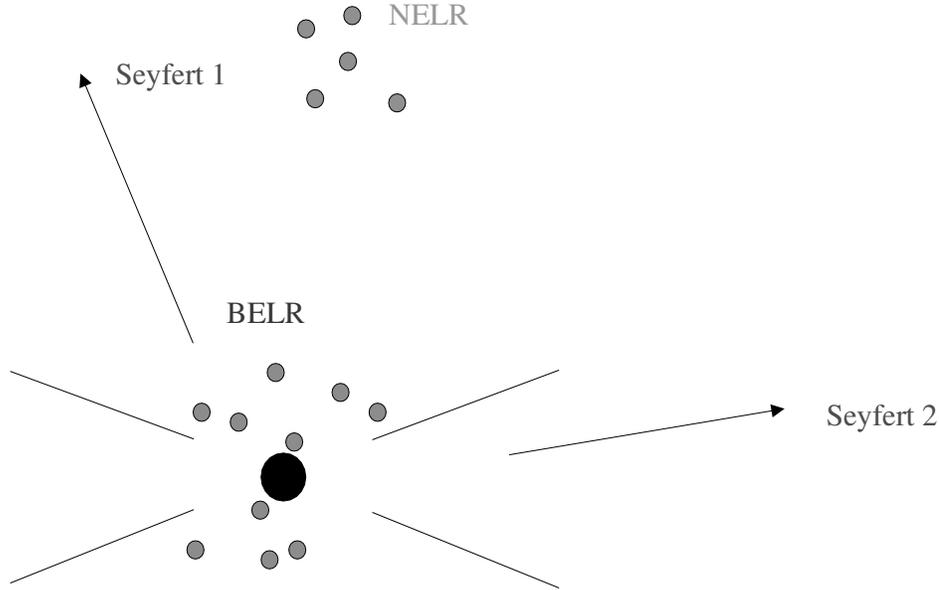}} \hfill
\parbox[b]{\hsize}{
\caption{Orientation model for Type 1 and Type 2 Sefert galaxies.  In
the simplest scheme, only the narrow line region  can be seen from a
Setfert 2 galaxy; the broad line region lies behind a thick equatorial
ring of obscuring gas and dust.}}
\end{figure}

Despite this, most authors have assumed that  the outflow of the
absorbing gas is roughly equatorial and radial, originating from close
to the hole. This implies that the quasar will only be classified as a
BALQ when the observer direction is also close to the equatorial
plane; otherwise she will see a regular radio-quiet quasar
(Fig.~(\thefigure)).  The goal is to use polarization
observations to see if this is truly the case (Antonucci 1993).

\subsection{Observation}
As we have remarked, a typical BALQ spectrum shows broad  emission
lines like CIV $\lambda1548$, accompanied by broad troughs extending
to the short wavelength end of the spectrum with widths up to
$\sim30,000$~km s$^{-1}$. The emission lines are generally
unpolarized, though both the semi-forbidden line  CIII]$\lambda1909$
and Ly$\alpha$ can be polarized, as we shall discuss. The troughs
which are caused by approaching gas, represent photons that are
removed from the radiation field and scattered sideways. They also
remind us that momentum  is taken out of the radiation field so that
the gas is accelerated (Arav \etal 1995). (Of course this may not be
the only accelerating force involved, though it is simplest  to assume
that it is.)

The absorption troughs themselves are not black which presumably means
that gas moving along different directions  is scattering radiation
into our line of sight. In addition, although the troughs are
relatively smooth, they do show velocity structure and a variable
degree of polarization that can be as large as $\sim0.2$ a
characteristic feature of resonance scattering.  However, resonance
scattering is not the only means of producing linear polarization;
electron scattering can do the job just as well and some observers
have preferred this explanation.

In what follows, we shall confine our attention to resonance
scattering for two reasons. Firstly, we observe resonance scattering
directly and its contribution to the opacity is three to four orders
of magnitude larger than that of electron scattering. Secondly, the
physics is much more interesting than that of electron scattering!
Furthermore we shall confine our attention to the polarization and the
implications that it has for the kinematics, as opposed to the
dynamics of the flow.

However, we must mention some additional observational clues as to the
nature of BALQs. Firstly, BALQs are both radio- and X-ray-quiet.  In
some models this implies that there is a highly ionized region that
can absorb the X-rays and transmit the ultraviolet radiation.  In
other interpretations, the X-rays are never emitted in the first
place. Secondly the BAL phenomenon is pretty much confined to quasars;
the lower power Seyfert galaxies do not exhibit these broad troughs.
Thirdly, the continuum (after subtracting the galactic contribution)
is fairly uniformly polarized with $p\sim0.01-0.05$,  suggesting
electron scattering in a disk  corona is responsible. This is
presumably located inside the  absorption line gas. As mentioned
above, it is also possible that this same scattering occurs outside
the absorption line gas in which case it would probably be responsible
for the polarization of  the absorption line troughs. Finally,  a
large polarization, increasing towards shorter  wavelength, has been
reported at wavelengths longward of the  Ly edge in a few, regular
quasars.  (It should be emphasized, though, that these HST
observations were very difficult to make and the results have been
controversial.)
\subsection{Physical Processes}
\subsubsection{Resonance Transitions}
Let us first consider the levels of intermediate Z ions adopting  the
Russell-Saunders approximation. We can distinguish the Li-like ions,
such as CIV, NV, OVI, the Be-like ions, such as CIII, NIV, OV and the
B-like ions such as CII, NIII, OIV. The electronic state is determined
by the principal quantum numbers for the valence electrons ($n,\ell$),
the total orbital and spin angular momenta ($L,S$) for the term and the
total angular momentum $J$ for the level which, in turn is divided
into $2J+1$ sublevels.  This information is encoded in the quantum
mechanical designation of  the level. For example, the ground state of
OIV is  $2s^22p\;^2P^o_{1/2}$ which means that one of the three
valence electrons  is unpaired in a $2p$ level with total orbital
angular momentum  $L=1$ (\ie a P state) and total spin $S=1/2$, hence
the superscript $2S+1=2$. There are two  possible choices for $J$ and
the lower energy one, by Hund's rule, has $J=1/2$ as designated in the
final subscript. (The final superscript, o,  indicates an odd parity,
which must change under a permitted (electric dipole)  transition.)
The electric dipole selection rules are that $\Delta \ell=\pm1;\, 
\Delta S=0;\, \Delta L=0,\pm1;\, \Delta J=0,\pm1$, (except that an
$J=0\rightarrow0$ transition is forbidden); and $\Delta M=0,\pm1$, (except
that a $M=0\rightarrow0$ transition is forbidden).
\subsubsection{Singlets}
Now consider a singlet transition such as the Be-like CIII$\lambda$977.
The ground state has $J=0$ and only one sublevel with $M=0$; the
excited state has $J=1$ and only permitted transitions to $M=\pm1$
need be considered. The radiation pattern depends only  upon the
angular parts of that wavefunction, through the Wigner-Eckart theorem.
When the scattering angle $\theta=\pi/2$,  the degree of linear
polarization can be computed to be  $p(\pi/2)=1$.  For a general
scattering angle,
\begin{equation}
p(\theta)={p(\pi/2)\sin^2\theta\over1+p(\pi/2)\cos^2\theta} \,,
\end{equation}
just as for Thomson scattering. This formula is generally true for
electric dipole transitions and so all we need compute is $p(\pi/2)$.
\subsubsection{Doublets}
\begin{figure}[t!]\label{fig_doublet}
\resizebox{\hsize}{!}{\includegraphics{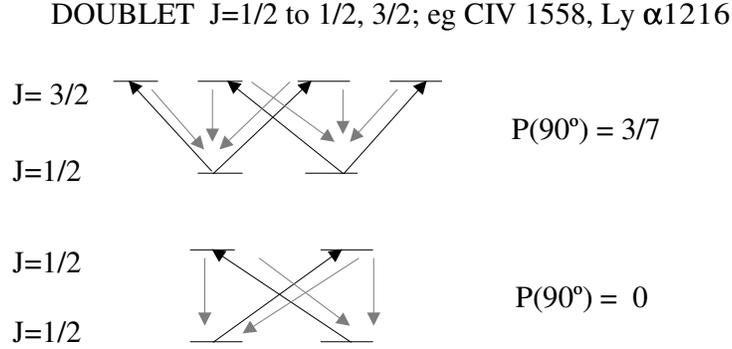}} \hfill
\parbox[b]{\hsize}{
\caption{Atomic transitions that contribute to the overall
polarization for a typical double transition. If the doublet is
resolved then the individual components have perpendicular
polarizations of $3/7$ and $0$.  If it is not, then the average
polarization is $3/11$}}
\end{figure}
The next most complicated case is the Li-like doublet transition, eg
CIV$\lambda1550$. Here the ground level has $J=1/2$ and there are two
possible excited levels of which the lower energy level is $J=1/2$.
All transitions between sublevels are permitted and the net
polarization is $p(\pi/2)=0$. The higher energy excited level,
associated with the shorter wavelength transition, has $J=3/2$ and
averaging over all of the permitted transitions gives
$p(\pi/2)=3/7$. If we average over both  transitions according to
their statistical weights, then we end up with $p(\pi/2)=3/11$
Fig.~(\thefigure).

$Ly\alpha$, has a similar type of transition.  Here the  the
wavelength separation of the two transitions is so small that the
doublet will not be resolved and   averaging over the two excited
levels makes sense.  However, in the case of CIV, the energy
difference is larger than the likely thermal width, so that one could,
for example, imagine continuum photons propagating out of an expanding
flow, encountering the polarizing $J=1/2\rightarrow3/2$ transition
first and then  encountering the $J=1/2\rightarrow1/2$ transition
which erases all of this polarization. It is clear that the
polarization is sensitive to the  nature of BALQ outflows.
\subsubsection{Triplets}
The next simplest case is the B-like triplet transitions such as CII
$\lambda1335$. In this case, there are two choices for the ground
state $J=1/2,3/2$ and two for the excited state, $J=3/2,5/2$; the
selection rules forbid  direct transitions with
$J=1/2\rightarrow5/2$. The energy difference between the two ground
levels is small enough that they should be equally populated  by
collisions in the absence of radiative transitions.

It is helpful to introduce the ionization parameter $U$ which  is the
ratio of the number density of hydrogen ionizing photons to the
electron density, designated $n$. When the  radiative excitation rate,
$\sim10^4Un_{10}$~s$^{-1}$, exceeds the collisional excitation rate
$\sim600n_{10}$~s$^{-1}$, the ground state sublevels will be populated
in an unequal fashion that must be computed by solving for all the
transitions. (Note that both of these rates are likely to be much less
than the spontaneous, de-excitation rate and so collisional
de-excitation of the excited states is generally thought  not to be an
issue for permitted lines under AGN conditions.)  Under these
conditions, when the ground sublevels are  radiatively mixed, the
resulting polarizations will differ from the values computed
assuming as statistical population of the ground sub-levels due to
collisions. In this case, the polarization is  increased from
$p(\pi/2)=0.21$ to $p(\pi/2)=0.38$.  This increase in predicted
polarization is typical.
\subsubsection{Supermultiplets}
The next level of complication arises when an ion in a single ground
state can be excited into several different excited states under
radiative mixing  conditions. In order to  solve for the population of
the different sublevels and the polarization,  we must consider these
distinct multiplets together.
\subsubsection{Magnetic Mixing}
A final variation, which is quite likely to be relevant in an AGN,
arises when the magnetic field is strong enough that the cyclotron
frequency, $\omega_G=1.8\times10^7(B/1{\rm G})$~rad s$^{-1}$ exceeds
the radiative  excitation rate. In this case, the relevant eigenstates
are referred to the magnetic  field direction rather than the normal
to the scattering plane and density matrices have to  be used to
attack the problem in general. This is also known as the Hanle effect
and  is discussed at greater length here in the solar context by
Dr. Stenflo.  An important consideration is the degree of Faraday
polarization.  The rotation of the plane of polarization is given by
\begin{equation}
\Delta\Phi\sim4\times10^{-5}\left({N\over10^{20}\,{\rm cm}^{-2}}\right)
\left({B\over1\,{\rm G}}\right)\left({\lambda\over1000\,\AA}\right)^2 \,.
\end{equation}
This is unlikely to be a factor at ultraviolet wavelengths, but could
be significant in the optical.
\subsection{Interpretation}
\subsubsection{Emission Line Clouds}
There is a standard model of the emission line gas based upon the
notion of an emission line cloud, a stratified slab of gas of size
$\sim10^{13}$~cm, located at a radius $R\sim0.3$~pc from the continuum
source  with a density $\sim10^{10}$~cm$^{-3}$. The ionization state
of the gas is determined by the relative importance of
photoionization and recombination which is, in turn, controlled by the
ionization parameter, $U$. Typically this is $U\sim0.1$.  These clouds
have a photoionization temperature $T\sim10,000$~K and an  equivalent
sound speed $\sim10$~km s$^{-1}$. However they are moving with speed
$\sim10,000$~km s$^{-1}$ and Mach number $M=1000$ through  a hotter
and more tenuous confining medium. This is patently absurd!
Nonetheless, this model does provide a good representation of the
ratios of the observed line strengths. What is  clearly required is a
convincing dynamical model that retains the  successful features of
the atomic astrophysics. For the moment, we just consider the
polarization in the context of the cloud model.

If we consider a CIV $\lambda1550$ photon propagating out of a
``standard'' cloud, the optical depth for solar abundance of carbon,
mostly in a triply ionized state, is $\sim10^5$. Under these
conditions, photons do not diffuse spatially out of the cloud as
might, at first, be guessed.  Instead, they undergo a random walk in
frequency and  escape when they migrate into the wings of the line
where the cloud becomes  transparent.  The polarization really has to
be computed using a Monte Carlo simulation and, under these conditions
of high optical depth,  it is not surprising that it is too small to
be measured, even when there is velocity  shear giving an anisotropic
escape probability.  The one conspicuous exception is the
semi-forbidden (intercombination line violating the selection rule
$\Delta S=0$)  CIII] line $\lambda1909$. Here the optical depth is
closer to $10$ and substantial linear polarization $p\sim0.01$, was
predicted and indeed has been reported in this line (Lee 1994,  Cohen
\etal 1995, but see Ogle \etal 1999).

There is a rather different story when external photons are scattered
by the emission  line clouds. Here we expect a high albedo and roughly
half the photons will undergo just one scattering (Korista \& Ferland
1998).  If the distribution of scatterers is anisotropic, then we
might also expect to detect a linear polarization signal in permitted
lines.  This is not usually seen, which suggests that a particular
line of sight  contains at most one cloud at a given wavelength.  This
is a strong constraint upon models of the velocity distribution.

\subsubsection{Rayleigh Scattering by Ly$\alpha$}
Hydrogen is the most abundant element so scattering of  Ly$\alpha$
photons is likely to occur far into the wings of the  line. The fine
structure level splitting  between $2P_{3/2}$ and $2P_{1/2}$ is quite
small for this transition and  off-resonance scattering is
characterized by the (classical) Rayleigh scattering phase function
(\eg Stenflo 1980).   This process is called Rayleigh scattering when
the initial and final states of the atom or ion are identical.(This
distinguishes  it from Raman scattering which arises when they are
different.)  Rayleigh scattering has been clearly observed in
symbiotic stars  (Nussbaumer, Schmid, \& Vogel   1989).  For large
velocity shifts $\Delta V$, the scattering optical  depth exceeds
unity for
\begin{equation}
{\Delta V\over 10^4\ {\rm km\ s^{-1}}}\simeq {\left[N_{HI}\over
3\times 10^{22}\ {\rm cm^{-2}}\right]^{1/2}} \,,
\end{equation}
(Lee \& Blandford 2000).

Of particular interest is the case of Ly$\alpha$, because in the
damping  wings the scattering phase function becomes that of the
classical Rayleigh function, which enhances polarization.  The
fundamental reason why this is the case is that near the line center,
where the optical depth is large, most photons migrate in frequency
space faster than they do in real space. They are therefore
comparatively insensitive to the cloud shape and large scale velocity
shear.  However, in the wings of the line, the optical depth is much
smaller and  a large scale pattern in the cloud shapes, for example,
translates into a measurable linear polarization. (Effects like this
have been  reported in high column density supershells associated with
starburst galaxies (Lee \& Ahn 1998).)

If the accretion disk of a quasar is warped so that some part  of it
is shadowed from the direct exposure to the central engine or has a
thickness that increases slower than linearly with radius, a large
column  density $N_{HI}$ may exist in the shaded region and Rayleigh
reflection  of Ly$\alpha$ is expected. In this case we may expect up
to 10 percent polarization in the Ly$\alpha$ wings,  which is
consistent with $\sim 7$ percent polarization reported from the
radio-quiet quasar PG~1630+377  (Koratkar \etal 1995)

\subsubsection{Absorption Line Clouds}
A somewhat analogous situation is found for the absorbing clouds
observed directly in the BALQs. These are believed to be located
outside the emission line region where the ionization parameter
$U\sim1$ and the size is  estimated to be even smaller than the size
of the emission line clouds $\sim(s/V)^2R\sim10^{11}$~cm.  (In a quite
different type of model, it has been proposed that the emitting and
absorbing gas originates from very much smaller radii and  forms part
of a space-filling flow, Murray \etal 1995. Many of the following
considerations apply to this model as well.)

The actual kinematics of line formation can be quite complicated.
This is generally handled under the Sobolev approximation (Rybicki \&
Hummer 1978).  An incident continuum photon is scattered when it is
resonant with a permitted transition taking place in the rest frame
of the outflowing gas. The scattered line is  redshifted in frequency
by $V_\parallel/\lambda$.  The surface occupied by gas resonant with a
fixed observer frequency is called a Sobolev surface and can have a
fairly convoluted shape.  Some photons may encounter several Sobolev
surfaces before escaping for good.  The optical depth to absorption by
ion $X$ through a single Sobolev surface  where the parallel velocity
varies monotonically is
\begin{equation}
\tau=\int ds \, n_X\sigma=\left|{dV_\parallel\over ds}\right|^{-1} \int
dV_\parallel n_X\sigma =\left|{dV_\parallel\over
ds}\right|^{-1}{c\over\nu}n_X \int d\nu \, \sigma(\nu) \,.
\end{equation}
Substituting numerical values,
\begin{equation}
\tau=0.3X_{-4}f \frac{\lambda}{1000\AA} \frac{dN_{20}}{dV_9} \,, 
\end{equation}
where $X=10^{-4}X_{-4}$ is the abundance of the ion, $f$ is the
oscillator strength  and $dN_{20}/dV_9$ is the hydrogen column density
(in units of $10^{20}$~cm$^{-2}$)  per unit velocity (in units of
10,000~km s$^{-1}$).  Resonance line scattering by the common ions
occurs at column densities about three orders of magnitude smaller
than those required for electron scattering (Lee \& Blandford 1997).

We can compute the polarization expected from a particular model using
Monte Carlo simulations. It turns out to be possible to give factor 2
estimates for the polarization, by multiplying expressions which
describe the most important factors in producing the integrated
polarization,
\begin{equation}
p\sim p(\pi/2) D(\tau) A G \,,
\end{equation}
In this equation, $D(\tau)$ is a depolarizing factor that takes into
account  multiple scattering. It is typically roughly fit by an
exponential $D=\exp(-\tau/b)$.  $A$ is a factor that takes into
account the anisotropy in the escape  probability. If the flow is
uniformly expanding, it will be very difficult for the photons  to
escape in the radial direction and far easier for them  to escape
tangentially. This roughly doubles the polarization. Conversely, if
the outflow is in the form of a jet, then small scattering angles will
be  favored with lower net polarization. Finally $G$ is  a
geometrical factor which is supposed to account for the observed gas
distribution.  If the outflow is confined to an equatorial fan
subtending a solid angle $\Delta\Omega$, then $G\sim\Delta\Omega/3$
and the electric vector will lie parallel to the projected  symmetry
axis.  For a jet, the polarization will be perpendicular to the axis
(which should be coincident with the radio axis).  More extensive
observations than have been possible so far will be needed to test the
hypothesis that the lines are due to resonance scattering, through the
dependence on  atomic type (through $p(\pi/2)$), and to decide upon the
flow geometry.  On this basis, the strongly polarized lines are
expected to be  HeI${\lambda584}$, OV${\lambda630}$, Ne
I${\lambda736}$,  NIV${\lambda765}$, CIII${\lambda977}$,
SiIII${\lambda1206}$, AlIII${\lambda1671}$ and
MgI${\lambda2852}$. The modestly polarized lines include
CII${\lambda687}$, OIV${\lambda789}$ and  NIII${\lambda991}$ and
weakly polarized lines include OVI${\lambda1034}$, HI${\lambda1216}$,
NV${\lambda1240}$,  SiIV${\lambda1396}$, CIV${\lambda1549}$ and
MgII${\lambda2798}$.  CII${\lambda858}$, NIII${\lambda764}$, and
OIV${\lambda609}$ are expected to be unpolarized  and can be used to
measure the amount of electron scattering.

\subsubsection{Polarization beyond the Lyman Edge} 
Observations of the rest ultraviolet continuum from a few high
redshift quasars have shown a strong polarization increasing
irregularly to shorter wavelengths  shortward of the Lyman continuum
(Koratkar \etal 1998). This may be as large as $p\sim0.2$, although
the  observations were extremely difficult and are consequently a bit
uncertain.  One possible explanation is that there are several highly
polarizing singlet lines in this region, like HeI$\lambda584$,
OV$\lambda630$, NeI$\lambda736$,  NIV$\lambda765$, or
CIII$\lambda977$. If the outflow speeds associated with these
relatively high ionization lines are large, $V\sim0.1c$, then it is
possible that the  lines could overlap enough to give an apparent
continuum polarization.
It would be good to have the capability to repeat these observations.

\subsection {Summary}
\begin{itemize}
\item Spectropolarimetry provides a powerful diagnostic of the disposition
of the broad emission and absorption line gas in quasars.
\item Resonance scattering should be variably polarized with the
degree and  direction dictated by fundamental considerations of atomic
astrophysics  and the flow geometry. The radiation observed in the
troughs  may be the scattered photons removed from other lines of
sight.
\item By contrast, the optical continuum exhibits a fairly  constant
polarization, suggestive of electron scattering.
\item Emission lines are generally unpolarized, excepting the
semi-forbidden line CIII]$\lambda1909$. This is consistent with the standard
cloud model of emission line formation.  BAL troughs are variably
polarized.
\item Large polarization rising with frequency has been reported to
the  blue of the Lyman continuum. This may be due to blends of
strongly polarized, prominent singlet transitions
\end{itemize}

\section{Neutron Stars}
\subsection{Motivation}
Although physicists and astronomers (most famously Baade and Zwicky)
were quick to appreciate the possibility that $\sim10^{57}$ neutrons
could assemble to form a self-gravitating neutron star, it was  not
until the discovery of radio pulsars in 1967, that there was
compelling evidence that they really existed. To date we have
cataloged over a thousand radio pulsars, know of hundreds of
accretion-powered neutron stars in X-ray binaries, and are starting to
find isolated neutron  stars accreting from the interstellar
medium. In addition, five  radio, or rotation-powered,  pulsars are
observed to pulse at optical wavelengths (Chakrabarty \& Kaspi 1998),
at least seven as $\gamma$-ray  pulsars (Thompson 2000), and $\sim40$
are detectable at X-ray energies (Becker, 2000).

For the astronomer neutron stars are the most common result of
evolution of a massive star.  However, far from being an endpoint,
they represent a rebirth often in a  more luminous state than the
progenitor star. For the physicist, neutron stars provide a
magnificent cosmic laboratory, allowing us to witness  the behavior of
cold nuclear matter at supranuclear densities, the indirect effects of
extremely high $T_C$ conductivity and superfluidity, and, as we shall see,
the consequences of magnetic field strengths perhaps nine orders of
magnitude greater than we can sustain on earth.

Neutron stars are also of special interest to the polarimetrist as
they have already furnished the strongest and most rewardingly
variable  signals of any cosmic sources. It is possible to follow the
change in the  polarization (sometimes nearly completely polarized)
through individual pulses from bright radio pulsars. It is also
possible  to study the average polarization properties of large
samples  of pulsars, viewed from a range of vantage points and,
thereby build  up a picture of the magnetic field geometry and try to
determine  the site of and the mechanism for their high brightness
emission.  Unfortunately, there is still no polarimetric capability at
X-ray wavelengths where accreting neutron stars in binary systems emit
most of their radiation.  However, very strong linear polarization is
anticipated and the details should be no less prescriptive of the
emission.

A third class of object, in addition to the accretion- and
rotation-powered pulsars that is of particular interest at the moment
is the magnetar. It appears that a minority of neutron stars are
formed with super-strong magnetic field  $\sim10^{14}-10^{15}$~G. As
predicted by Thomson and Duncan (1995), these magnetars decelerate
very quickly but still have a larger  reservoir of magnetic energy
that can be tapped to power $\gamma$-ray bursts. These field
strengths are well in excess of the quantum electrodynamical critical
field $B_c\equiv m_e^2c^3/e\hbar= 4.4\times10^{13}$~G, where the
cyclotron energy of an electron  equals its rest mass.  This, in
principle, allows us to test the theory in a regime that is
qualitatively quite different from that in which impressively  high
precision tests have already been made. (There is no real anxiety
that the theory is suspect above the critical field but, as is the
case with general relativity, there is a strong interest in performing
the check.)  Another class of X-ray source is the Anomalous X-ray
Pulsars.  These are possibly a late evolutionary phase of magnetars.

In this section, we shall discuss the expected polarimetric properties of all
three types of sources, although there are only observations of
rotation-powered pulsars.
\subsection{Observation}
\subsubsection{Rotation-Powered Pulsars}
Radio pulsars are spinning, magnetised neutron stars (\eg Lyne \& Smith
1998). The majority have  spin period between 0.1s and 3s and surface
magnetic field strengths $\sim10^{12}$~G, estimated from the rate at
which they appear to  slow down. The neutron stars themselves appear
to be mostly formed with masses quite close to the Chandrasekhar mass
$\sim1.4$~M$_\odot$.  Their poorly measured radii are $\sim10$~km,
consistent with there  having central densities a few times nuclear as
the best models of  the nuclear equation of state imply. (We really do
not know the  interior composition at all well. It could be mostly
neutrons or  contain a large fraction of protons, hyperons, pions or
even  free quarks. Accurate measurements of the radius along with the
rate of cooling will provide important constraints on the equation of
state of  nuclear matter.) Other, impressive vital statistics of
neutron stars include escape velocities $\sim0.3c$, surface gravities
$\sim10^{14}$~cm s$^{-2}$ and maximum spin frequencies (that are
nearly  attained in observed objects) $\sim1$~kHz.  The radio emission
from pulsars has extremely high brightness temperatures which can, by
some estimates,  exceed $\sim10^{30}$~K.

The integrated pulse profiles of radio pulsars frequently show  one,
two or three pulses. This, and other observational evidence, has been
interpreted in terms of an emission model where there is a strong
``core'' beam  of emission close to the magnetic axis surrounded by a
weaker ``cone'' beam.  When the observer latitude is similar to that
of the magnetic axis, a single, dominant core component is
seen. Increasing (or decreasing)  the observer latitude leads to a
three peaked, cone-core-cone pattern.  When the observer is more
inclined to the magnetic axis, only the two  cone components will be
seen. These pulse profiles exhibit strong,  broad band  linear
polarization varying through the pulse. Values $p\sim1$ are consistently
measured at certain  pulse phases in certain pulsars. The position
angle swings regularly through the main pulse with a total swing that
can be as high as  $\sim180^\circ$.  Strong circular polarization is
also commonly measured near the center of the pulse, with the
handedness often changing sign.

Individual pulses, whose polarimetric properties can be measured in
the  strongest pulsars, are no less interesting. They show individual
emission units, known as subpulses, with durations typically a few
degrees of pulsational phase. These can ``drift'' through the pulses
appearing at progressively earlier or later phases in successive
pulses. These  subpulses often exist in one of two orthogonal
polarization states.  Even shorter timescale features known as
microstructure (or now even  nanostructure) has been well documented
and these too can exhibit high, though complex, polarization
properties.

The optical pulses from the Crab pulsar in the Crab Nebula have been
particularly well studied. The pulse profile is cusp-like and the
plane of polarization swings smoothly through $\sim70^\circ$, while
the degree varies between $p=0.1$ and 0.5. Another famous optical
pulsar is associated with the Vela supernova remnant, shows similar
strong, variable linear polarization.
\subsection{Physical Processes}
\subsubsection{Curvature Radiation}
The large measured brightness temperatures  imply that the radio
emission is produced  by a coherent process. One widely discussed
possibility  is that the emission is some variant on coherent
curvature emission whereby  bunches of charged particles stream
outward along the curving, roughly  dipolar magnetic field lines from
the star with ultrarelativistic speed (Lorentz factors $\gamma$  of
several hundred) and radiate like giant electrons.  The emission
properties are like those already summarized for  synchrotron
radiation, with the important difference that the radius of curvature
of the orbit, $R$, is energy-independent. The characteristic emission
frequency is $\omega\sim\gamma^3c/R$, lying in the radio band for
$R\sim10-100$~km. The radiation from an individual bunch is beamed
within an angle $\sim\gamma^{-1}$ to the direction of motion.

At a particular pulse phase, the observer sees emission from a curve
though the magnetosphere where the line of sight is tangent to the
magnetic field. There is thought to be a radius-to-frequency mapping
so that the emission at a given frequency is concentrated over an
interval of radius along this curve, and that this radius decreases
with increasing frequency. The polarization  from a tangent point will
be quite strongly linearly polarized with electric vector parallel to the
projected curvature vector on the sky.  As the pulsar  spins, this
projected curvature vector will rotate on the sky and a
characteristic swing of the plane of polarization will be produced.
This is known as the rotating vector model.  If we view an individual
bunch from one side of its orbital plane then the other, we will see
one  sense of circular polarization followed by the opposite
sense. This mechanism clearly has the ingredients to explain the radio
polarization  observations. However, it is a bit puzzling how a
totally polarized pulse can be  formed in this manner. One possible
explanation is that the bunches are quite strongly flattened and they
radiate most strongly perpendicular to their flattening plane, where
the polarization will be most strongly linear.
\subsubsection{Maser Processes}
There has recently been a resurgence of interest in maser emission
models. It is relatively easy to imagine that the necessary
population inversion will develop in the outflowing plasma. A typical
pulsar can develop an EMF of $\sim10^{13}-10^{16}$~V and  a small
fraction of this potential difference developing in a transient
``gap'' will create a fast stream of electrons and/or positrons that
can stream through more slowly moving (though still ultrarelativistic)
plasma. What is a bit harder is to find suitable  wave-particle
interactions that can lead to an overall negative absorption
coefficient. As an illustration of the difficulty, note that
synchrotron  radiation {\it in vacuo} has an absorption coefficient
\begin{equation}
\kappa={1\over8\pi\nu^2}\int
dN{1\over\gamma^2}{\partial\over\partial\gamma}
\gamma^2p_\nu(\nu,\gamma) \,.
\end{equation}
The single particle emissivity, $p_\nu(\nu,\gamma)$ is proportional to 
$\nu^{1/3}\gamma^{-2/3}$ at low frequency and increases with energy at 
high frequency.  Hence, synchrotron absorption is
necessarily positive and maser action is precluded, independent of the
particle distribution function. Similar conclusions have been drawn
for other emission mechanisms in vacuum.

However this conclusion does not necessarily follow if there is a
plasma present. One particularly interesting case is the  so-called
anomalous cyclotron (otherwise known as cyclotron-Cerenkov) resonance
between a wave with angular frequency $\omega$ and wave vector $\vec
k$ interacting with an electron (or positron) moving with velocity
$\vec v$, in a field where the non-relativistic gyro frequency is
$\omega_G$. This occurs if
\begin{equation}
\omega-k_\parallel v_\parallel=-\omega_G/\gamma \,.
\label{resonance}
\end{equation}
where $\parallel$ refers to the component along the magnetostatic
field.

This equation needs some interpretation (Lyutikov, Blandford \&
Machabeli 1999). Consider, for simplicity, a circular polarized wave
propagating along the magnetic field.  A resonance satisfying the
condition Eq.~(\ref{resonance}) clearly requires that the  phase
velocity of the wave be less than $c$.  If we transform into the
guiding-center frame of the electron,  then the wave angular frequency
changes sign indicating that it is propagating in the  opposite
direction along the magnetic field, with the same sense of circular
polarization. This means that it resonates with particles gyrating in
the opposite sense around the field (i.e. with opposite charge) than
is the case in a regular cyclotron resonance. Put another way the
particle outruns the wave so that, in the frame where the wave is at
rest, the particle follows the electric vector as it spirals around
the magnetic field.  A consequence is that, in the electron guiding-center 
frame, a quantum of wave energy has  negative energy. Therefore
in exciting the electron to a higher state of gyration it emits a
quantum and {\it vice versa}. (In the rest frame, population 
inversion now requires having more  particles in the lower gyrational
state!) Not surprisingly, this arrangement can lead to maser action in
the outer magnetosphere and this has been proposed as a mechanism to 
produce the core
emission. The modes are naturally  circular polarized, although the
handedness depends upon the details of the electron and positron
distribution functions.

\begin{figure}[t!]\label{fig_pulsar}
\resizebox{\hsize}{!}{\includegraphics{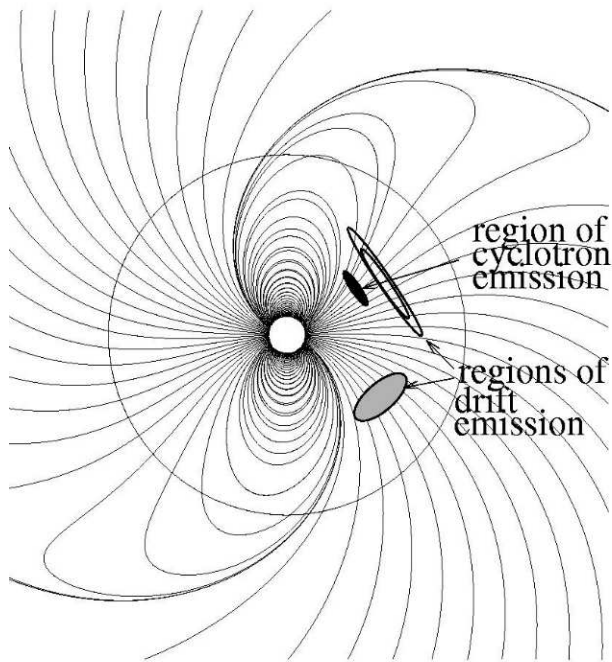}} \hfill
\parbox[b]{\hsize}{
\caption{Location of core and conal emission regions in a maser  model
of pulsar emission (after Lyutikov \etal 1999).}}
\end{figure}

This is not the only possible way to have a maser process.  There is a
second resonance associated with the curvature drift motion $v_{{\rm
drift}}$ of the gyrating electron as it moves along the curving
magnetic field
\begin{equation}
\omega-k_\parallel v_\parallel=k_\perp v_{{\rm drift} \,.}
\end{equation}
This will be perpendicular to the curvature plane and will
consequently produce emission with polarization  orthogonal to that
predicted by the rotating vector model. This mechanism  has been
invoked to account for the cone emission (Fig.~(\thefigure)).

\subsubsection{Propagation Effects}
There is unfortunately a complication (Arons \& Barnard 1986, Lyutikov
\etal 1999, Hirano \& Gwinn 2001).  The emitted radiation must
propagate  through the outer magnetosphere. This can lead to genuine
absorption  at the normal cyclotron resonance by more slowly moving
electrons.  Landau damping is also a possibility. Furthermore, there
are refractive effects that may imprint additional polarization  on
the emergent radiation in much the same way as occurs in the
ionosphere.  There can be mode conversion, for example from a
subluminal ordinary  mode to a propagating electromagnetic wave.
Finally, and perhaps most interestingly from a physics perspective,
there can be non-linear scattering effects (Lyutikov, 1998). At the
high brightness temperature (or, equivalently, large occupancies of
individual quantum mechanical  states) found in pulsar radiation there
will be a large amount of scattering between different radiation
beams. These  can be mediated by individual electrons, in which case
the interaction is known as induced Compton scattering, or by
collective wave modes of the plasma (known as  Raman scattering when
the scatterer is an electrostatic wave). There is not space to discuss
these rather complex processes further, save to remark that, the
associated matrix elements have a strong sensitivity to polarization
and frequency and that this can be used  to identify them. The theory
is starting to match the observations in its richness!

Some of these propagation effects can also be relevant in the
interstellar medium.
\subsubsection{Thomson Scattering in a Strong Magnetic Field}
The cross section for Thomson scattering must be changed if there is a
strong magnetic field present (\eg M\'esz\'aros 1992).  In the limit
when the wave angular frequency $\omega<<\omega_G$, or
$E<<44(B/10^{12}{\rm G})$~keV, the electrons are constrained to move
along the magnetic  field like beads on a wire. The dominant cross
section is between polarization states with $\vec k,\vec E, \vec B$
coplanar.  It is clearly given by
\begin{equation}
{d\sigma\over d\Omega}=r_e^2\sin^2\theta\sin^2\theta' \,,
\label{magcs}
\end{equation}
where $\theta,\theta'$ are the angles made by the incident and
scattered wave vectors with the magnetostatic field.
\subsubsection{Inverse Compton Scattering in a Strong Magnetic Field}
This anisotropy in the cross section introduces an additional
complication to inverse Compton radiation.  This is because the
incident photon propagates in a direction making an angle
$\sim\gamma^{-1}$ to the magnetic field and so the total scattering
cross section as given by Eq.~\ref{magcs} is reduced by a factor
$\sim\gamma^{-2}$ from the Thomson value assuming that the frequency
in the electron rest frame ($\omega'$) is less than
$\omega_G$. However, under these circumstances, we must  also consider
the effect of the ``$\vec E\times\vec B$'' drift of the  electron
perpendicular to the magnetic field. This will produce an oscillatory
electron motion that is larger than the field parallel motion by a
factor $\gamma\omega'/\omega_G$.  The azimuth and incident
polarization-averaged cross section will be given by
\begin{equation}
{d\sigma\over d\Omega}={r_e^2\over4}\left({\omega\over
\omega_G}\right)^2(1+\cos^2\theta)(1+\cos^2\theta') \,,
\end{equation}
for incident frequency $\omega$ satisfying
$\omega_G/\gamma\omega<\omega_G$. In practice, this leads to a rather
complex polarization pattern.

\subsubsection{Quantum Electrodynamical Effects}

The virtual electron-positron plasma that comprises the QED vacuum
affects the propagation of radiation through it.  These are clearly
likely to be of importance when the magnetic field strength
approaches the critical field strength, $B_c$.

An external  magnetic field causes photons of different polarizations
to travel at  slightly different speeds.  The $CP$-invariance of
electrodynamics tells us  that the two modes must be linearly
polarized.  Specifically, the indices of  refraction of both modes
differ from unity and are given by (Heyl \& Hernquist 1997a):
\begin{eqnarray}
n_\perp &=& 1 +  \frac{\alpha}{4\pi} \frac{8}{45} \left (
\frac{B_\perp}{B_c} \right )^2 + \cdots \,,\\ n_\|    &=&  1 +
\frac{\alpha}{4\pi} \frac{14}{45} \left ( \frac{B_\perp}{B_c} \right
)^2 + \cdots \,,
\end{eqnarray}
for $\hbar \omega \ll\ m_e c^2$ and $B \ll\ B_c$.  $B_\perp$ is the
component  of the magnetic field perpendicular to the propagation
direction of the  photon.  A photon in the perpendicular polarization
has its electric field  vector perpendicular to the projection of the
magnetic field into  the transverse plane, and similarly for the
parallel polarization.

For fields stronger than  $B_c$, the index of refraction for the mode
with the electric field  perpendicular to the external magnetic field
saturates at
\begin{equation}
n_\perp = 1 + \frac{\alpha}{4\pi} \frac{2}{3} \sin^2\theta + \cdots \,,
\end{equation}
while the index for the other mode increases without limit
\begin{equation}
n_\|  = 1 + \frac{\alpha}{4\pi} \frac{2}{3} \sin^2\theta \frac{B}{B_c}
 + \cdots \,.
\end{equation}
Therefore, there are two natural limits to the behavior.  In the weak
field regime the vacuum polarization may be sufficiently strong to
decouple the polarization states as they propagate through the
magnetosphere (Heyl \& Shaviv 2000, 2001).  Normal pulsars as well as
strongly magnetized white dwarfs fall in this regime.  In the
strong-field regime magnification and distortion of the image of the
neutron star surface may become important in addition to the
decoupling of the propagating modes (Shaviv, Heyl \& Lithwick 1999).

The processes of photon splitting and one-photon pair production are
forbidden in field-free regions, the first by Furry's theorem and the
second by four-momentum conservation.  However, in the strong magnetic
fields surrounding a neutron star, both processes may be important.
Photon splitting most strongly affects photons in the $\perp$-mode
which may split into two photons in the $\|$-mode (Adler 1971, the
mode-naming convention used here is opposite to that used by Adler);
this both distorts and polarizes the photon spectrum.

One-photon pair production has an energy threshold of  $\hbar \omega
\sin \theta > 2 m_e c^2$ for photons in the $\|$-mode.  The threshold
for photons in the $\perp$-mode is slightly larger $\hbar \omega \sin
\theta >  m_e c^2 ( 1 + \sqrt{ 1 + 2 B/B_c} )$.  Near the thresholds,
in strong magnetic field especially, the cross-section for this
process is complicated by the formation of the pair in discrete
Landau levels or a positronium bound state (Daugherty \& Harding 1983;
Usov \& Melrose 1996).

\subsection{Interpretation}

\subsubsection{Rotation-powered Pulsars}
Partly because it provides such a straightforward interpretation  of
the polarization data, the curvature  radiation model is probably the
favorite explanation for pulsar radio emission. However, it does have
some drawbacks. One of these is that it is difficult  to maintain a
compact bunch for very long as the electrons travel along trajectories
with different radii of curvature and are subject to radiation
reaction. Another problem is that it has proven  hard to find a
suitable plasma instability which will allow  charge particle bunches
to grow. A third and currently controversial  observational claim is
that the emitting area is much larger than expected if the bunches
form in the inner  magnetosphere as the rotating vector model
requires. Specifically, Gwinn, \etal (2000) find that the size of the
Vela pulsar emission region is roughly ten per cent of its light
cylinder radius, $\sim c/\Omega$, suggesting that the emission
originates in the outer magnetosphere.  Conversely, Cordes (2001) finds
that the source is unresolved.

The maser explanation, by contrast, only works in the  outer
magnetosphere and the average direction of polarization  should be
orthogonal to the projection of the pulsar spin axis on the sky.  This
can be tested using the  Crab and Vela pulsars where X-ray jets are
observed which, although their  formation is not understood, are
presumed to be along the projected spin axis.  In the case of the Crab
pulsar, the situation is ambiguous because it is  unclear if the two
pulses come from one or two magnetic poles. However, in the case of
the Vela pulsar, the electric vector is unambiguously  perpendicular
to the projected spin axis, consistent with the maser model.  It could
also be consistent with curvature radiation if propagation  effects
are important. It should be possible to discriminate between these two
models observationally.

The pulsed optical radiation and X-ray radiation seen from several
radio  pulsars is generally thought to be incoherent synchrotron and
inverse Compton radiation, respectively.  However, the location of the
emission region, and, in the case of the X-rays, the source of the
incident photons (the surface of the star or coherent radio emission
from the magnetosphere) is currently undecided. Suffice it to remark
it here that polarization  arguments figure prominently in these
debates.

\subsubsection{Accretion-powered Pulsars}
In addition to being strongly polarizing, the opacity,
Eq.~\ref{magcs},  is highly anisotropic. So, even though we cannot
measure the X-ray polarization directly, at present, it does have a
strong effect on what we observe.  In particular, the mass that
accretes onto a spinning neutron star with a surface field
$\sim10^{12}$~G is likely to be channeled toward the magnetic poles
and, if the accretion rate is large enough, there will be a
significant Thomson  opacity at the poles. However, it will be much
easier for the radiation to escape along the direction of the magnetic
field than in the transverse direction. For this reason, X-rays are
thought to emerge in two broad pencil beams about the magnetic axis as
is observed.
\subsubsection{Surface Emission from Isolated Neutron Stars and Magnetars}

The atmospheres of neutron stars are thought to emit strongly
polarized radiation (Pavlov \& Shibanov 1978).  The opacities in the
two polarization modes of the atmospheric plasma may differ by several
orders of magnitude (Lodenqual \etal 1974).  The opacity in the
extraordinary mode (\ie $E \perp B$) is generally a factor of
$(\omega/\omega_G)^2$ smaller than in the ordinary mode.  Since the
atmospheres are typically at a temperature of several million degrees,
the natural place to study this emission is in the X-rays.
Furthermore, as we shall see, the vacuum significantly affects the
propagation of radiation passing through it at X-ray and higher
energies.

Although the emission at the surface may nearly be fully polarized,
one observes radiation from regions with various magnetic field
directions.  In this vein, Pavlov \& Zavlin (2000) argue that the net
polarization in the X-rays is on the order of ten percent and
decreases for more compact stars.   However, this treatment ignores
the fact that QED renders the vacuum birefringent.  The field strength
varies sufficiently gradually, that is
\begin{equation}
\frac { \nabla |\Delta k| }{|\Delta k|} \ll\ |\Delta k| \,,
\end{equation}
where
\begin{equation}
|\Delta k| = \frac{\alpha}{4 \pi} \frac{2}{15} \frac{\omega}{c} \left
 ( \frac{B_\perp}{B_k} \right )^2 \,,
\end{equation}
in the weak field regime, that the two polarization modes are
decoupled.  Radiation produced at the surface with its polarization
direction perpendicular to the local magnetic field direction will
keep its polarization perpendicular to the field even as it passed
through regions where the field direction changes.  The observed
polarization reflects the direction of the field at a distance
\begin{equation}
r \approx 1.2 \times 10^{7} \left ( \frac{\mu}{10^{30}~{\rm G cm}^3}
\right )^{2/5} \left ( \frac{\nu}{10^{17}~{\rm Hz}} \right)^{1/5}
\left ( \sin \beta \right)^{2/5} {\rm cm} \,,
\end{equation}
from the center of the star (Heyl \& Shaviv 2000, 2001).  Here $\mu$
is the magnetic dipole moment of the neutron star, and $\beta$ is the
angle between the dipole axis and the line of sight.  QED ensures that
the strongly polarized radiation at the surface of the neutron star
remains strongly polarized until it is detected; therefore, the simple
detection of strongly polarized X-rays from the atmosphere of a
neutron star will verify a thus far untested prediction of QED.  Cheng
\& Ruderman (1979) used a similar argument to account for the strong
polarization of radio emission from pulsars.

Although, the QED process of one-photon pair production plays a
crucial role in radio pulsars by fueling the plasma that produces  the
emission ({\it e.g.} Daugherty \& Harding, 1982), the threshold
for the reaction is much higher than the typical energies from the
surface emission.  The cross-section for photon splitting increases
dramatically with increasing photon energy, $\propto E^6$ (Heyl \&
Hernquist 1997b) and is only important above 10~keV even in the
strongest magnetized sources.

\subsection{Soft-Gamma Repeaters}

Thompson and Duncan first argued that the soft-gamma repeaters are
neutron stars fueled by a dynamic magnetic field whose strength
greatly exceeds $B_c$.  In their quiescent state, these objects emit
thermal radiation from their surfaces and the discussion of the
previous subsection applies.  If their surface fields are sufficiently
strong (the surface field is expected to exceed the value estimated by
spin down of $\sim 10^{15}$~G), magnetic lensing may be important for
photons whose polarization is parallel to the magnetic field ({\it
i.e.} the ordinary mode); however, thermal emission in this mode
appears to be strongly suppressed.

However, what makes the soft-gamma repeaters unique is that they
burst.  In fact, the soft-gamma repeater, SGR 1900+14, is the only
object beyond our solar system to have had contemporary geophysical
consequences (it ionized the nightside upper atmosphere nearly to
daytime levels).  This soft gamma-ray emission is generally well below
the threshold for one-photon pair production, but photon splitting
should degrade the energies of the photons by at least a factor of two
and polarize them by converting photons in the extraordinary mode to
the ordinary mode (Baring \& Harding 1997).  Observing this tracer of
photon splitting would require gamma-ray polarimetry.

\subsection {Summary}
\begin{itemize}
\item 
Magnetized neutron stars provide cosmic laboratories where we can
observe unique polarization effects in action and use them to identify
the emission mechanism.
\item 
Radio pulsars offer the richest polarization data set outside the
solar system. They are strongly diagnostic of the emission mechanism
and the effects of propagation.
\item 
Accretion-powered pulsars introduce new effects associated  with strong
field anisotropic emission and scattering in the $\sim10^{12}$~G
surface fields.  Even though the X-ray polarization is not yet
measured, it is  important in determining the total spectrum and pulse
profile.
\item X-ray pulsars allow us to address important physics questions,
like the composition and compressibility of cold matter at
supra-nuclear density.
\item The simple detection of strongly polarized X-rays from the atmosphere
of a neutron star (or optical radiation from neutron stars with 
$B > 10^{13}$~G) will verify the prediction that QED renders the
vacuum birefringent and provide an estimate of the radius of the star itself.
\item 
The convincing case that magnetars exist with surface fields well in
excess of the critical field, ($4.4\times10^{13}$~G), offers the
equally exciting, (though observationally very challenging) prospect
of testing quantum electrodynamics in a regime far removed from
terrestrial investigation.
\end{itemize}

\section{Black Holes}
\subsection{Motivation}
There is now very good evidence for the existence of black holes in
the universe.  They appear to be a common endpoint of the evolution of
massive stars in  our Galaxy and nearby galaxies and we know of
roughly ten good cases where the dynamically determined mass
significantly exceeds the Oppenheimer-Volkoff limit (for neutron
stars) $\sim2.5$~M$_\odot$ and the Chandrasekhar limit (for white
dwarfs) $\sim1.4$~M$_\odot$. In addition, there are many more cases of
transient X-ray sources where the circumstantial evidence, in the
absence of dynamics, is pretty convincing. A significant fraction of
massive stars must end their life this way.

Similarly, dynamical studies of the nuclei of nearby galaxies reveal
the  presence of ``massive dark objects'' which, if they were, for
example, clusters of compact objects, would be very
short-lived. Identifying them with massive black holes is by far the
most conservative conclusion to draw. It appears that the nuclei of
most normal galaxies, including our own, contain black holes with
masses in the range $\sim3\times10^6-3\times10^9$~M$_\odot$.  (There
are speculative  suggestions that there may be a large population of
intermediate mass black holes, perhaps relics of the first generation
of  stars.)

The existence of black holes is, arguably, the most far-reaching
implication of the general theory of relativity. The theory has been
probed in the weak field regime and passed all quantitative tests with
an accuracy that can be as small as $\sim3\times10^{-4}$.  It is in
the nature of the theory that,  if we understand the laws of physics
under these circumstances, it is simply a question of geometry to
describe strong field environments, when the equivalent Newtonian
potential approaches $c^2$. If the theory in its essential simplicity
is correct, then the metric of an asymptotically flat black hole
spacetime (excluding some mathematical niceties) is essentially
given. Indeed, in one of the greatest successes of mathematical
physics, we have a closed form version of the metric of a spinning
black hole, known as the Kerr metric, and essentially all classical
physics that can be discussed in a flat spacetime can also be
discussed around a black hole; there are no difficulties of
principle. There are, however, considerable difficulties in execution
(which have mostly been overcome using numerical  calculations.)
Although we know of more general spacetimes that include  a
gravitationally significant charge or orbiting mass, we believe that
these are irrelevant to observed black holes and that astronomers need
only  be concerned with the Kerr metric.

However, it is logically possible that the theory of general
relativity  could be wrong or incomplete and that, as a consequence,
black holes  are fundamentally and observably different from their
general relativistic description.  For this reason, it is vitally
important that we try to find ways to probe the spacetime around black
holes, now that we know where to find them. In this regard, observing black
holes  provides a far more telling test of relativity theory than
cosmological  observations. This is because cosmological observations
are seriously compromised by our deep ignorance of the nature of dark
matter  and energy as well as the effects of evolution.

We already know that gravitational waves exist. Binary pulsars  are
observed to lose orbital energy at rates that agree with theory to a
fraction of a percent. However, this mostly tests linearized theory
even in the sources.  The direct detection of gravitational  waves,
which we hope will happen one day, is also is only a linear test.
The ultimate test of general relativity is to make detailed
observations of gravitational waves from coalescing black holes; an
observation that  I suspect will be not be technologically feasible
for some time.
In addition to testing strong field relativity, this  can also provide
much useful astrophysical information on  galaxy merger rates, AGN
evolution and so on. Computing the wave forms in  necessary generality
is a major challenge to computational science. What is relevant in the
present context is that  gravitational waves have natural polarization
states, just like electromagnetic waves, and much of the  information
from these coalescences will be encoded in the polarization
details. (There are other strong field sources of gravitational
radiation that have been considered, notably topological defects like
cosmic strings. Unlike the case with  black holes, there is no
observational evidence yet for their existence. However, if they are
ever discovered, then it may well be their gravitational radiation
polarization that is their distinctive signature. Computing this
polarization is a good project which appears to have been mostly
ignored.)

There is a second and quite different reason for being interested in
black holes.  This is that we do not understand properly how they
work. We have already introduced jets, disks, and outflows; and discussed
how polarization observations can teach us about their properties. We
already know, from direct observation, that all three of these
continue down to relatively close (in logarithmic terms) to the black
hole. This is where most of the energy is  released. However, we do
not understand how this all happens and how the  flow around the black
hole depends upon the mass and the spin of the hole  as well as the
accretion rate and the immediate environment. It appears that the
answers to these questions will only be found by  exploring the black
hole itself.
   
A third contextual aspect of this study is that  we are beginning to
suspect that black holes have a much  larger and more active role in
galactic and extragalactic astronomy than used to be the case.  It is
increasingly likely that gamma ray bursts are associated with the
formation or augmentation of black holes and  that these have major
environmental impacts on their  surroundings and could soon become
useful cosmological probes. Black hole  transients provide the
dominant hard X-ray emission of galaxies like our own  and create
powerful outflows. The discovery of dormant, or near-dormant  black
holes in the nuclei of normal galaxies has affirmed the long-standing
black hole model of active galactic nuclei, including quasars and
giant, double radio sources. However, the implications have much
broader implications than the properties of AGN {\it per se}. The
quasars themselves provide the best cosmologically distant beacons
that we have and they allow us to study the intergalactic medium,
matter and cosmography. Furthermore, it is becoming increasingly
apparent that they have an much more active role in the very formation
of galaxies, both in the initiation and perhaps in the cessation of
the process.

The long term observational goal, then, is to  verify that black holes
are described by the  Kerr metric and to measure their masses and
spins in such a way as to elucidate  their role in stellar and
galactic evolution. In this section, we will try to show  how
polarization observations can contribute to  meeting this objective.
\subsection{Observation}
We have already described most of the relevant observations of black
holes including the fairly strong dynamical measurements of their
masses and the  Fe K line emission which provides the strongest
evidence to date that black holes spin relatively rapidly. The most
relevant, existing observation for the purpose of this lecture are the
measurements of linear and circular polarization.
\subsubsection{Sgr A$^\ast$}
The center of our Galaxy appears to be identified with the radio
source  Sgr A$^\ast$. There is now excellent dynamical evidence that
is a ``dark, compact object'' with a mass $2.6\times10^6$~M$_\odot$
and a black hole  is by far the most conservative interpretation. The
source is nearly at rest and  stars can be tracked moving (and
accelerating) around it (Ghez \etal 2000).  The source has a spectrum
which peaks at $\sim300$~GHz and the image is broadened at radio
wavelengths by interstellar scattering. At 43~GHz, the average
scatter-broadened size is reported to be $\sim3\times10^{13}$~cm, (Lo
\etal 1999) and to scale roughly $\propto\lambda^2$.  Sgr A$^\ast$ is
a weak and soft X-ray source.  Interestingly, a 106 d periodicity in
the radio emission has also been reported (Zhao, Bower \& Goss 2001).

Sgr A$^\ast$ has long been known to have negligible linear
polarization at  radio and mm wavelengths. This is not a surprise
because the Faraday rotation is expected to be quite high so that the
differential (in both angle  and frequency) rotation is also large
enough  to depolarize all measurements. What is a surprise (as
discussed here by Hildebrand) is that  Aitken \etal (2000) measure
$\sim10$~percent linear polarization at 150~GHz, using the SCUBA
instrument on the James Clerk Maxwell Telescope, although there is an
upper limit of $\sim1$~percent at 86~GHz (Bower \etal 2000).  Clearly
there is a need to confirm the SCUBA measurement.

More recently, it has been discovered that Sgr A$^\ast$ exhibits quite
strong circular polarization, (Bower, Falcke \& Backer 2000). The
5~GHz degree of circular polarization appears to have been stable at a
value of $\sim-0.003$ for nearly twenty years. At higher frequencies,
up to $\sim43$~GHz, the degree of circular polarization appears to
increase up to a few percent and become increasingly variable,
doubling in a few days.
\subsection{Physical Processes}
\subsubsection{Spinning Black Holes}
The spacetime around a spinning black hole is described by the Kerr
metric expressed in Boyer-Lindquist coordinates with $G=c=1$
\begin{eqnarray}
ds^2=&-&(1-2mr/\rho^2)dt^2-
(4amr\sin^2\theta/\rho^2)dtd\phi+(\rho^2/\Delta)dr^2+\rho^2d\theta^2\nonumber\\
&+&(r^2+a^2+2mra^2\sin^2\theta/\rho^2)\sin^2\theta d\phi^2 \,,
\end{eqnarray}
where
\begin{eqnarray}
\rho^2&=&r^2+a^2\cos^2\theta \,,\\ \Delta&=&r^2-2mr+a^2 \,,
\end{eqnarray}
and $m$ is the mass such as would be measured by the orbit of a
distant  satellite, and $a<m$ is the specific angular momentum of the
hole, as  could be measured operationally by the precession  rate of a
gyroscope.

There is an event horizon, ${\cal H}$, which is located where
$\Delta=0$ \ie where the radial coordinate $r=r_+=m+(m^2-a^2)^{1\over
2}$.  Particles on timelike or null geodesics must be inwardly moving
within $r_+$ which leads to the interpretation that ${\cal H}$ represents 
a surface of no return.  The four velocity, $\vec u
=\{dt/d\tau,dr/d\tau,d\theta/d\tau,d\phi/d\tau\}$ of a material
particle satisfies
\begin{equation}
g_{\alpha\beta}u^\alpha u^\beta=-1 \,.
\end{equation}
The equation of a photon, following a null geodesic  is given by
$g_{\alpha,\beta}dx^\alpha dx^\beta=0$, supplemented with equations
representing the conservation of energy and angular momentum as well
as an additional integral of the motion.

The angular velocity $\Omega=d\phi/dt$ of a particle, orbiting  with
fixed $r,\theta$, therefore satisfies
\begin{equation}
u^{02}[g_{00}+2\Omega g_{0\phi}+\Omega^2g_{\phi\phi}]=-1 \,.
\end{equation}
This implies that  $\Omega_{{\rm min}}<\Omega<\Omega_{{\rm max}}$ 
where $\Omega_{{\rm min}}>0$ when 
$r_+<r<r_e\equiv m+(m^2-a^2\cos^2\theta)^{1/2}$. The radius  $r_e$ is known 
as the static limit and the region between it and the horizon, where inertial
frames are dragged by the spin of the hole, is known as the
ergosphere. A particular significance of the ergosphere is that orbits
of negative energy (including rest mass) exist within it.   As
$r\rightarrow r_+$ at the event horizon,
\begin{equation}
\Omega_{{\rm min}}\rightarrow\Omega_{{\rm max}}
\rightarrow\Omega_H\equiv a/(r_+^2+a^2)^{1/2} \,,
\end{equation}
the angular velocity of the hole.

A remarkable theorem due to Hawking states that the area of the
horizon, which can be computed from the metric to be $A=\int_{{\cal
H}}(g_{\theta\theta}g_{\phi\phi})^{1/2} d\theta d\phi=4\pi
(r_+^2+a^2)$ cannot decrease. We can use this to define a so-called
irreducible radius $r_0$ and irreducible mass $m_0$ through
\begin{equation}
r_0=2m_0=(A/4\pi)^{1/2} \,.
\end{equation}
This immediately implies that $a=r_0^2\Omega_H$. It turns out that the
area is proportional to the thermodynamic entropy. Now imagine that we
exchange  some mass and some angular momentum with the hole,
reversibly (and therefore at constant area) from just outside the
horizon. These must be added according to
\begin{equation}
dm=\Omega d(am) \,.
\end{equation}
This equation can be integrated to give
\begin{equation}
m={m_0\over[1-(\Omega r_0)^2]^{1/2}} \,.
\end{equation}
Imposing the condition $a<m$, we find that there is a mass
$m-m_0<0.29m$ that can, in principle, be extracted from the hole.  The
main way that this is thought to occur naturally is through the agency
of large scale magnetic field that threads the event horizon of the
black hole. This magnetic field can exert a torque on the hole,
similar to the magnetic torques acting upon the sun and neutron stars,
for example.   Black hole spin provides a plausible power source for
high energy phenomena like ultrarelativistic jets and gamma ray bursts
and this is one reason why black holes are commonly thought to be
spinning rapidly. (Even if the  spin is not a significant power
source, then the specific angular momentum of the gas that accretes
onto a black hole is generally so large that it is very hard to
imagine slowly spinning holes ever being formed.)

For present purposes, though, what is most important  is that, in a
rapidly spinning hole, the accreting matter can form a disk  extending
quite close to the horizon. Specifically, if we consider circular
Keplerian orbits around a hole then these are stable down to a radius
of marginal stability which is located at $6m$ for a non-rotating
(Schwarzschild) hole and approaches the horizon as $a\rightarrow
m$. In addition, it is possible for strong pressure gradients within
the  disk to support matter in non-Keplerian orbits inside $6m$.  This
means that gas may survive quite a long while in and around the
ergosphere before crossing the horizon or being ejected. The dominant
emission may come from this region, and as the radiation escapes, its
trajectory and the  propagation of its polarization can be significantly
influenced by the curvature of the spacetime. This provides us with a
potential probe of the Kerr metric.
   
\subsubsection{Geometrical Optics of Plasma Waves in Flat Space}
We are interested in the propagation of plasma waves in the curved
spacetime around a black hole. However, for the moment, let us
consider the propagation of waves in flat space under geometrical
optics. This is appropriate because the wavelengths that we
are considering, at least for electromagnetic radiation, are always much
smaller than the horizon radius.  It is convenient to exploit the
analogy with Hamiltonian  particle dynamics. Under the eikonal
approximation, we can define  a phase $\phi$ such that
$\nabla\phi=\vec k$ and $\partial\phi/\partial t=-\omega$.  The phase
velocity is, as usual, defined by $\vec V_\phi=\omega/\vec k$.  We
assume the existence of a dispersion relation
\begin{equation}
\omega=\Omega(\vec k,\vec x,t) \,.
\end{equation}
Equivalently, there is a Hamilton-Jacobi equation
\begin{equation}
{\partial\phi\over\partial t}+\Omega(\nabla\phi,\vec x,t)=0 \,,
\end{equation}
which must be satisfied. The three Hamilton equations are
\begin{eqnarray}
\label{dkdt}
{d\vec k\over dt}&=&-\nabla\Omega \,,\\ {d\vec x\over
dt}&=&{\partial\omega\over\partial\vec k}\equiv\vec V_g \,,\\
{d\omega\over dt}&=&{\partial\Omega\over\partial t} \,,
\end{eqnarray}
where
\begin{equation}
{d\over dt}\equiv{\partial\over\partial t}+\vec V_g\cdot\nabla \,,
\end{equation}
and $V_g$ is recognized as the group velocity.  These three equations
govern the propagation of plasma modes in a spatially inhomogeneous
and temporally varying medium, under the short wavelength
approximation. In a cold, unmagnetized plasma the dispersion relation
is $\Omega=(\omega_P^2+c^2k^2)^{1/2}$ and $V_g=c^2/V_\phi$. We can
think of wave quanta -- plasmons -- and the energy they carry, as
moving along a path $\vec x(t)$ with the group velocity. These
plasmons are conserved; \ie the wave energy density $U$
can  be shown to obey a conservation equation of the form
\begin{equation}
{\partial\over\partial t}\left({U\over\omega}\right)+\nabla\cdot
\left({U\vec V_g\over\omega}\right)=0 \,.
\label{intensity_cons}
\end{equation}

Consider, for example, shear Alfv\'en waves.  The
dispersion relation is $\omega=\vec k\cdot\vec B/(4\pi\rho)^{1/2}$ 
and the group velocity is $\vec V_g=\vec B/(4\pi\rho)^{1/2}$. The wave 
packets propagate along the magnetic field along with the energy 
although $\vec k$ can be directed at a large angle to $\vec B$.

The propagation of the polarization can be most simply approached by
decomposing the given wave into its normal modes, propagating
each along the direction of the group velocity, and compute the
relative change in phase (to lowest order in the eikonal approximation
this is  simply $\int dx\cdot\vec k$ along the path.)  Of course the
character (\ie polarization, local phase velocity \etc) of these modes
will change, but provided we are in the WKB limit, the modes are
distinguished, non-degenerate (see below), and there is no mode
crossing (which can occur and has to be handled more carefully), 
these changes will change adiabatically.  As a result, they can be 
tracked and the total phase difference along a path can be computed.



The total flux can be computed by using the conservation of intensity
along the path.  Equivalently, we say that the phase space density of
individual quanta of wave excitation, in individual modes is conserved
along paths.  If there is emission or absorption along the path then
it is straightforward to write down the equation of radiative transfer
and use the local emission and absorption coefficients to evolve
the intensity (\eg Rybicki \& Lightman 1979; Bekefi 1966).


In practice, of course, all of this can easily become  quite
involved. However it is important to understand the principles
because these alert us to the sort of effects we might expect to
observe.
\subsubsection{Magnetized Accretion Disk} 
As a more pertinent illustration of some of these ideas, let us
consider electromagnetic wave modes propagating through an accretion
disk containing a strong, though disordered, magnetic field.

Consider a magnetoactive plasma with $X=\omega_P^2/\omega^2,
Y=\omega_G/\omega<1$ where $\omega_P$ is the plasma frequency and
$\omega_G$ is the electron gyro frequency.  Under so-called
``quasi-longitudinal''  conditions - essentially when $\cos\theta<Y$,
where $\theta$ is the angle between $\vec k$ and  $\vec B$, the
electromagnetic eigenmodes  are elliptically polarized with axis ratio
$r=1\pm Y\sin\theta\tan\theta$  and phase velocity difference $\Delta
V=cXY\cos\theta$.

Now suppose that  synchrotron (or cyclotron) radiation is emitted
within  an accretion disk of thickness $H$. The major axis of the
polarization ellipse will be Faraday rotated at a rate
$\Delta\Phi/ds=\Delta V\omega/2c^2$.  Now if, as we expect, the
magnetic field direction reverses often along a ray and if, as also
anticipated,$X|Y|\omega H/c>>1$, then we expect that the  emergent
linear polarization will be vanishingly small (cf \S1.3.4).  (The
limiting polarization along an individual ray is likely to be
determined by the decrease in the density rather than the magnetic
field strength.)

The circular polarization is a bit more problematic.  If we suppose
that there is a net magnetic field normal to the disk, as is true of
some  models, then there should be a  preferred sense of the circular
polarization that is emitted in cyclotron or low energy synchrotron
radiation.  This will be largely preserved in propagating out of the
disk.

If the emitted radiation is effectively unpolarized, we can analyze
the production of circular polarization due to Faraday conversion by
decomposing each wave into the eigenmodes.  Thus consider a  single
eigenmode propagating out of the disk through a spatially  varying
magnetic field.   We suppose that the variation happens relatively
slowly on the scale of the  wavelength so that the polarization
ellipse adjusts adiabatically  (with no mode crossings).  Next,
suppose that the two eigenmodes are launched with equal amplitude and
that the field is uniform. The beating between the  two eigenmodes
will result in a circular polarization of amplitude
$Y\sin\theta\tan\theta$  that changes sign as the plane of linear
polarization rotates.  If either the Faraday depth is large, or the
sign of the magnetic field is as likely to be negative as positive,
then the limiting circular  polarization that  emerges from the disk
is equally likely to have either sign and so there will be no net
circular polarization.


Now let the magnetic field direction vary along a ray.  If the angle
$\theta$ varies, then there will be a corresponding change in the axis
ratio of the polarization ellipse, but still no preference for one
sign over the other. However, if the azimuthal angle  $\phi$ relating
$\vec B$ to $\vec k$ changes  in a systematic fashion along all rays
then a net phase difference between the two modes will develop.  In the
same way that Faraday conversion creates circular polarization, such
a phase difference will also create circular polarization.  The main 
difference is that while Faraday conversion depends upon the direction of the
magnetic field, and hence will not lead to a net polarization for randomized
fields, the new mechanism depends upon the rate of shearing and only 
the strength of the field.
Thus, it is possible to conceive of situations in which the rate of shearing
and the typical length scales over which the magnetic field reverses
are related in such a manner that a net circular polarization is produced 
without a commensurate linear polarization. 


This situation is precisely what might be anticipated in a magnetized 
accretion disk. In the disk interior, the typical field direction will 
trail to reflect the differential rotation in the disk. However, the field
will be swept back by progressively smaller angles as the ray
approaches the disk surface, corresponding to a net rotation of the
average azimuthal angle $\phi$. There will only be a preferred sense
of limiting circular polarization, if the  magnetostatic field is
still changing in this systematic manner over the last radian of
Faraday rotation. The net circular polarization will be $\sim c\ln
Y/\omega HX$.

These are some of the subtle effects that could be present  in an
accretion disk and which could, under some circumstances, create
measurable polarization, even in the absence of general relativity.
\subsubsection{Geometrical Optics of Vacuum Waves in a Curved Spacetime} 
We first consider the propagation of photons in a vacuum surrounding a
black hole. These follow orbits called null geodesics, just as
material particles follow timelike geodesics. The equation of motion
can be  expressed in a general coordinate system though, in our case,
Boyer-Lindquist coordinates, by saying that the total derivative of
the wave vector along the ray vanishes. In index notation, this becomes
\begin{equation}
\label{geoeq}
k^\beta k^\alpha_{;\beta}=0 \,,
\end{equation}
cf Eq.~(\ref{dkdt}).

There are essentially three constants of the motion that describe
these orbits, an energy, $-k_0$ an angular momentum, $k_\phi$, and a
third quantity known as the Carter constant, $Q$ (\eg Misner, Thorne \&
Wheeler 1973.) (In fact, we only need the ratios  $k_\phi/k_0, Q/k_0$
to define the rays.)  Close to the black hole the rays are strongly
curved  with the consequence that a distant observer, able to resolve
a black  hole, would be able to see a distorted image of the disk
behind the hole, apparently hovering above the hole (cf
Fig.~(2).). The ray trajectories  are given, in
general, by the solution of a set of coupled ordinary  differential
equations that can be partly integrated in terms of  elliptic
functions (Rauch \& Blandford, 1994). Given a model of the emission,
for example of the surface emissivity of a thin accretion disk, it is
a straightforward, though quite lengthy, exercise to compute the total
emergent flux and, indeed, the form of the image that would  be
resolved if the black hole could be resolved.

Now, turn to the propagation of the polarization of vacuum modes in a
curved spacetime, specifically outside the horizon of a Kerr hole
(Laor, Netzer \& Piran 1990). As we have already emphasised, both
non-thermal emission (\eg synchrotron radiation) and electron
scattering are likely to create polarized sources of radiation. If we
just consider linear  polarization for the moment, and the
generalization to circular polarization is straightforward, then the
question that we must answer is ``How do  we propagate the plane of
polarization from one point to the  next along a curving ray?''. The
answer is that the electric vector is  ``parallel-transported''(eg
Misner, Thorne \& Wheeler 1973).  What this means is that the unit
vector in the direction of the electric field $\hat e^\alpha$ changes
along the ray such that its magnitude and projection onto the direction 
of the ray remain constant.  An additional constraint arising from Maxwell's
equations in vacuum is that the electric vector must be perpendicular to the
wave vector.  Hence, in index notation, we have
\begin{equation}
\label{ppe}
k^\beta\hat e^\alpha_{;\beta};\quad\hat e^\alpha k_\alpha=0 \,.
\end{equation}
These can be solved consistently to propagate the electric vector along
a ray.

What is actually done is somewhat different. It turns out that there
is another conserved quantity, called the Walker-Penrose (1970)
tensor, associated with the photon spinors (the familiar geometrical
object  that in this case are associated with light-like geodesics,
the path taken  by photons in vacuum.)  This actually involves the
electric vector and can  be used to relate the polarization  at the
point of emission to that at the point of observation directly without
having to integrate a differential equation (Connors, Stark \& Piran
1980).  It is then possible to define a transfer function for the
polarization and to compute the polarized flux given a specific
emission model using the propagated intensity. (Note that, in
propagating the intensity, we must correct for the  Doppler and
gravitational shifts. There is a natural way to do this in general
relativity.)
\subsubsection{Geometrical Phase}
The next level of complication is to introduce the plasma into the
curved  space time.  Let us do this in two stages. The first stage is
to  ignore the magnetic field so that the local  dispersion relation
takes the form $\Omega=(\omega_p^2(\vec x)+c^2k^2)^{1/2}$.  In this
case, the refractive index is locally isotropic.  This means that the
two eigenmodes at a point are degenerate and that we have to formulate
a rule to connect the polarization from one point along a path to the
next.

The wave packets, which travel at the local group velocity, are no
longer moving along null geodesics, but timelike geodesics instead.
It turns out that the Hamiltonian equations of motion can be
generalized in a covariant manner, provided that one has knowledge of
the local linearized dispersion relation at every relevant point in
space time.  Therefore, there is a prescription for computing the
paths. These can be  described by a four velocity $u^\alpha$ and an
acceleration  $a^\alpha=du^\alpha/d\tau$ with respect to a
freely-falling frame, where $d\tau$ is an interval of proper time.
The standard relativistic way to handle this is to generalize the
notion of parallel-transport to Fermi-Walker transport (\eg Misner,
Thorne \& Wheeler 1973), which corrects for the non-null motion of the
wave packets. The propagation equation becomes
\begin{equation}
\label{fwt}
{d\hat e^\alpha\over d\tau}=\hat e^\beta a_\beta u^\alpha- u_\beta
\hat e^\beta a^\alpha \,.
\end{equation}
Eq.~(\ref{fwt}) reduces to Eq.~(\ref{ppe}) when Eq.~(\ref{geoeq}) is
satisfied.  This provides a natural basis in which to discuss
polarization  propagation and phase changes.

It is instructive to consider a wave propagating along a  twisting
optical fiber, with  $\vec k$ parallel to the local tangent to the
fiber.  Here again we have gradients in an isotropic refractive
index. In this case, the unit electric vector, along $\hat{\vec e}$,
must remain perpendicular  to the unit wave vector $\hat{\vec
k}$. When the fiber bends, the change in  $\hat{\vec e}$ must be along
$\hat{\vec k}$; there is no other vector to be involved  as the medium
is isotropic.  Therefore we can write down the equation of propagation
for the electric  vector from first principles.
\begin{equation}
{d\hat{\vec e}\over ds}=-\hat{\vec k}\left(\hat{\vec
e}\cdot{d\hat{\vec k}\over ds}\right) \,.
\end{equation}
This is a limiting case of Eq.~(\ref{fwt}).

A good way to visualize what is happening (Berry 1990) is to allow the
tangent to the fiber to trace out a path on the unit sphere.
$\hat{\vec e}$ is tangent to the sphere and it is straightforward to
see that rotation angle of $\hat{\vec e}$ after traversing a complete
circuit equals the solid angle enclosed by that circuit. If we
propagate a linearly polarized wave along a twisting fiber, the
polarization direction will, in general, be rotated between two points
where the fiber is  parallel.  This experiment has been performed
successfully (Chiao \etal 1989). (Actually  this was under conditions
when physical as opposed to geometrical  optics applies, though the
results should be identical.)

This rotation -- essentially a phase change between the two circularly
polarized eigenmodes -- is known as the geometric phase. Geometric
phase is  a quite general phenomenon in physics and analogs are
expected to be  relevant to wave propagation in a curved
spacetime. The Foucault pendulum provides another example of this
general phenomenon. As is well known, a Foucault pendulum at latitude
$\ell$ will only rotate through an angle in  inertial space of
$2\pi(1-\sin\ell)$, the solid angle traced out by the radial vector on
the unit sphere, as it is carried around a complete circuit in one day
by the spinning Earth. Now to see where general relativity may come
in, it is helpful to consider a Foucault pendulum at the North pole.
According to the above discussion, there is no rotation of the plane
of oscillation according to Newtonian dynamics.  However, the tiny
dragging of inertial frames effect associated with the spin of the
earth leads to an equally tiny rotation of the plane of polarization
and there have been  proposals to measure it. Effects such as these
would be much larger near  a spinning black hole and could also
influence the propagation of  electromagnetic waves.
  
\subsubsection{General Relativistic Magnetoionic Theory}
The second stage is to reinstate the magnetic field which, on general
grounds, is surely present. This breaks the degeneracy between the two
wave modes. This is akin to changing the pivot of a Foucault pendulum
from a point attachment to an axle. As far as is known,  there is no
generalization  of the Carter constant and the Walker-Penrose tensor,
to non-null wave modes.  The equations for the paths can be integrated
and the intensity  and the polarization can be propagated using the
relativistic generalization  of the approach outlined above (Broderick
\& Blandford, in preparation.)

When a linearly-polarized vacuum wave crosses the ergosphere of a
rapidly spinning hole, there  is a contribution to the emergent
polarization position angle of order unity due to the dragging of
inertial frames. We can think of this as a phase difference in of
order unity in the two circular polarized modes  into which the linear
mode can be decomposed. Now, if we introduce a  magnetoactive plasma
into the path, then there is likely to be a large  Faraday rotation
per unit length. The total rotation will differ by  much more than
$O(1)$ along different paths that any emitted linear polarization is
likely to be erased.  However, as discussed above, the eigenmodes are
not completely circular  and have an ellipticity $O(XY)$. What this
means is that a systematic phase  difference $O(1)$ will be introduced
between the two modes and that, if the  original modes are in phase so
that there is no circular polarization, a degree of polarization
$O(XY)$ will emerge, independent of the reversals  of the magnetic
field and variations in the total Faraday  rotation along different
lines of sight.
 
These matters deserve further attention.

\subsection{Interpretation}
\subsubsection{Sgr A$^\ast$ and other Low Luminosity AGN}
X-ray observations of Sgr A$^\ast$ (Baganoff \etal 2001) have shown
the the luminosity is very small ($\sim4\times10^{33}$~erg s$^{-1}$)
and the spectrum is quite soft. Variability on an hour timescale may
also have been seen. (As has been argued elsewhere, Blandford \&
Begelman 1999, this is generally to be expected if most of the mass
supplied to the hole is blown away in a wind.)  This suggests that the
density of gas is very low close to the hole and opens  up the
possibility  that we may be seeing radio or, more likely, mm emission
from  the ergosphere. Under these conditions, polarization
observations could be quite diagnostic of  the physical conditions.

The variable circular polarization discussed above, increasing in degree with
frequency, has at least three explanations. Firstly,
there could be a radius to frequency mapping so that the radio
photosphere shrinks with frequency and the field gets stronger  so
that the energy of the emitting relativistic electrons also decreases.
This leads to an increase in the emitted degree of circular
polarization (cf \S1). Variability studies at high frequency should be
quite  diagnostic. Secondly, the polarization could be due to a flat
space propagation effect along the lines discussed above. It will be
particularly interesting to see if the sign of the circular
polarization  really does not change, as the observations to date may
suggest.  This could, in principle, be related to the angular velocity
of the  disk as outlined above, though quantitatively this seems
improbable.

The third possibility may be the most  unlikely, yet it is the most
exciting. This is that the circular polarization reflect directly the
geometry of the ergosphere and be due to a general relativistic,
propagation effect as outlined above. In the case of Sgr A$^\ast$, we
expect the gas at high latitude in the ergosphere to be moving with
speed $\sim c$.  We deduce that the expected degree of circular
polarization is:
\begin{equation}
C\sim XY\sim10^{-2}\left({\dot M\over10^{22}\,{\rm g\,s^{-1}}}\right)
\left(B\over100\,{\rm G}\right) \left({\lambda\over1\,{\rm cm}}\right)^3 \,.
\end{equation}
It is not impossible that this  effect is observable  These matter
deserve more attention, both observational and theoretical, in Sgr
A$^\ast$ and other, nearby galaxies.

\subsubsection{Imaging the Ergosphere}
Interesting and timely as these ideas may be, radio observations are
probably not likely to contribute to a confirmation of the essential
features of the Kerr metric until we can actually image the
ergosphere.  At present, as we have remarked, the best resolution has
been achieved  in M87 ($\sim100m$). Probably the best prospects lie
with sub mm VLBI observations of Sgr A$^\ast$, where interstellar
scattering precludes resolving the ergosphere at radio wavelengths
(Falcke, Melia \& Agol 1999).  There are also quite futuristic plans
to develop X-ray interferometry to achieve analogous goals (Cash \etal
2000).

\subsection {Summary}
\begin{itemize}
\item Black holes are common features of the evolution of massive stars.
\item Massive black holes are commonly found in the nuclei of normal
galaxies.  Presumably they powered active galactic nuclei including
quasars and  radio sources in the past.
\item We have good grounds to be confident in the general theory of
relativity and, specifically, the Kerr metric which describes the
curved spacetime around a spinning black hole. However, this does not
absolve us from the  responsibility of testing the theory.
\item We also want to understand how black holes accrete and how they
form jets as well as their impact on Galactic and extragalactic
astronomy.
\item Recent observations suggest that we may be observing radio and
mm  emission from very close to the black hole in Sgr A$^\ast$.  There
is consequently interest in developing the magnetoionic theory  and
radiative transfer in a general relativistic environment.
\end{itemize}
\begin{acknowledgments}
RB thanks Javier Trujillo Bueno and the Director of the Instituto de
Astrofisica de Canarias for their hospitality,  his fellow lecturers
for their instruction and the students for their attention and
questions. The NSF and NASA are acknowleged for support under grants
AST 99-00866 and  5-2837 respectively.
\end{acknowledgments}

\end{document}